\documentclass[epj
]{svpreprint} 
\usepackage{graphicx}
\usepackage{amsmath}
\usepackage{amssymb}
\begin{document}
\titlerunning{Operation Regimes and Slower-is-Faster-Effect in Traffic Control} 
\title{Operation Regimes and Slower-is-Faster-Effect in the Control of Traffic Intersections}
\author{Dirk Helbing and Amin Mazloumian
}                     
%
%
\institute{ETH Zurich, UNO D11, Universit\"atstr. 41, 8092 Zurich, Switzerland}
\date{Received: date / Revised version: date}
%
\abstract{The efficiency of traffic flows in urban areas is known to crucially depend on signal operation.
Here, elements of signal control are discussed, based on the minimization of overall travel times or vehicle queues. Interestingly, we find different operation regimes, some of which involve a ``slower-is-faster effect'', where a delayed switching reduces the average travel times. These operation regimes characterize different ways of organizing traffic flows in urban road networks. Besides the optimize-one-phase approach, we discuss  the procedure and advantages of optimizing multiple phases as well.  To improve the service of vehicle platoons and support the self-organization of ``green waves'', it is proposed to consider the price of stopping newly arriving vehicles.
} 
\maketitle
\section{Introduction}

The study of urban traffic flows has attracted the interest of physicists for quite a while (see, e.g., Refs. \cite{bml,nagacity,esser1997,nagel1}). This includes the issue of traffic light control and the resulting dynamics of vehicle flows \cite{Chowdhury1999,Brockfeld2001,Fouladvand2001,Sasaki2003,Toledo2004,Nagatani2006,Piccoli}.
Theoretical investigations in this direction have primarily focussed on single intersections and grid-like street networks, e.g. adaptive control \cite{Barlovic2004,Fouladvand2004a,Fouladvand2004} of a single traffic light or coordination of traffic lights in Manhattan-like road networks with unidirectional roads and periodic boundary conditions. Some of the fascination for traffic light control is due to the relationship with the synchronization of oscillators \cite{Syn1,Sync1,Laemmersync} and other concepts of self-organization \cite{Nakatsuji1995,Sekiyama2001,Helbing2005a,Gershenson2005,Gershenson2007,Helbing2005,Helbing2007,jstat}.
\par
The efficiency of traffic light control is essential to avoid or at least delay the collapse of traffic flows in traffic
networks, particularly in urban areas. It is also crucial for attempts to reduce the fuel consumption and CO$_2$ emissions of vehicles. Both, delay times and acceleration maneuvers (i.e. the number of stops faced by vehicles)\footnote{For formulas to estimate these quantities as a function of the utilization of the service capacity of roads see Ref. \cite{EPJB8}.} cause additional fuel consumption and additional CO$_2$ emissions \cite{PhDLaemmer}.
Within the USA alone, the cost of congestion per year is estimated to be 63.1 billion US\$, related with 3.7 billion hours of delays and 8.7 billion liters of ``wasted'' fuel \cite{Schrank2005}. Climate change and political goals to reduce CO$_2$ emissions force us to rethink the design and operation of traffic systems, which contributes about one third to the energy consumption of industrialized countries. On freeways, traffic flows may eventually be improved by automated, locally coordinated driving, based on new sensor technologies and intervehicle communication \cite{Schoen,Kest1}. 
\par
But what are options for urban areas? There, traffic lights are used to resolve conflicts of intersecting traffic streams. In this way, they avoid accidents and improve the throughput at moderate or high traffic volumes.
For a discussion of the related traffic engineering literature, including the discussion of traffic light coordination and adaptive signal control, see Refs. \cite{Helbing2005,jstat} and references therein.
In the following, we will focus our attention on some surprising aspects of traffic flow optimization.

\subsection{Paradoxical Behavior of Transport Systems}

Besides Braess' paradox (which is related to selfish routing) \cite{braess,braess2,anar}, the slower-is-faster effect is another counter-intuitive effect that seems to occur in many transport networks. It has been found for pedestrian crowds, where a rush of people may delay evacuation \cite{nAture}. 
\par
Slower-is-faster effects have fascinated scientists for a long time. Smeed \cite{Smeed}, for example, discussed "some circumstances in which vehicles will reach their destinations earlier by starting later'', but Ben-Akiva and de Palma \cite{Moshe} showed that this effect disappeared under realistic assumptions. Moreover, it is known from queuing theory that idle time can decrease the work in process (i.e. basically the queue length) in cyclically operated production systems under certain circumstances, particularly when the variance in the setup times is large \cite{Cooper}. These circumstances, however, do not seem to be very relevant for traffic light control.
Nevertheless, there are many examples of slower-is-faster effects in traffic, production, and logistic systems, and it has been suggested that the phenomenon is widespread in networked systems with conflicting flows that are competing for prioritization \cite{pAtent,produc}. While there are numerical algorithms to exploit this effect systematically to improve the performance of these systems \cite{pAtent}, there have been only a few analytical studies of the slower-is-faster effect \cite{oscianal,meanfie,uli}. Therefore, we will put a particular focus on the study of conditions leading to this counterintuitive, but practically relevant effect.
\par\medskip
Our paper is structured as follows: While Sec. \ref{speci} specifies the traffic system investigated in this paper, Sec. \ref{flowopt} discusses the throughput of intersections. Section \ref{trati} continues with the problem of minimizing travel times, while Sec. \ref{opq} discusses the minimization of queue lengths. The challenge in these sections is to come up with a concept that still leads to reasonably simple formulas, allowing one to study the behavior of the proposed signal control analytically. A successful approach in this respect is the ``optimize-one-phase approach'', which seems justified by the short intervals, over which traffic flows can be anticipated reliably. Among the operation regimes resulting from the optimization process are also some with extended green times, corresponding to a
``slower-is-faster effect'' (see Sec. \ref{peropreg}). A further improvement of signal operation is reached by applying multi-phase optimization, when flow constraints are taken into account. As Sec. \ref{MUL} shows, this approach leads to a variety of plausible operation regimes. A summary and dicussion is presented in Sec. \ref{sUm}. Complementary, Sec. \ref{plato} will discuss the ``price'' of stopping vehicles, which is an interesting concept to support moving vehicle platoons (and, thereby, the self-organization of ``green waves''). For a more sophisticated, but analytically less accessible approach to the self-organization of coordinated traffic lights and vehicle streams in road networks see Refs. \cite{jstat,pAtent,anticip}.

\section{Specification of the Traffic System under Consideration}\label{speci}

\begin{figure}[htbp]
\begin{center}
\includegraphics[width=6cm,angle=180]{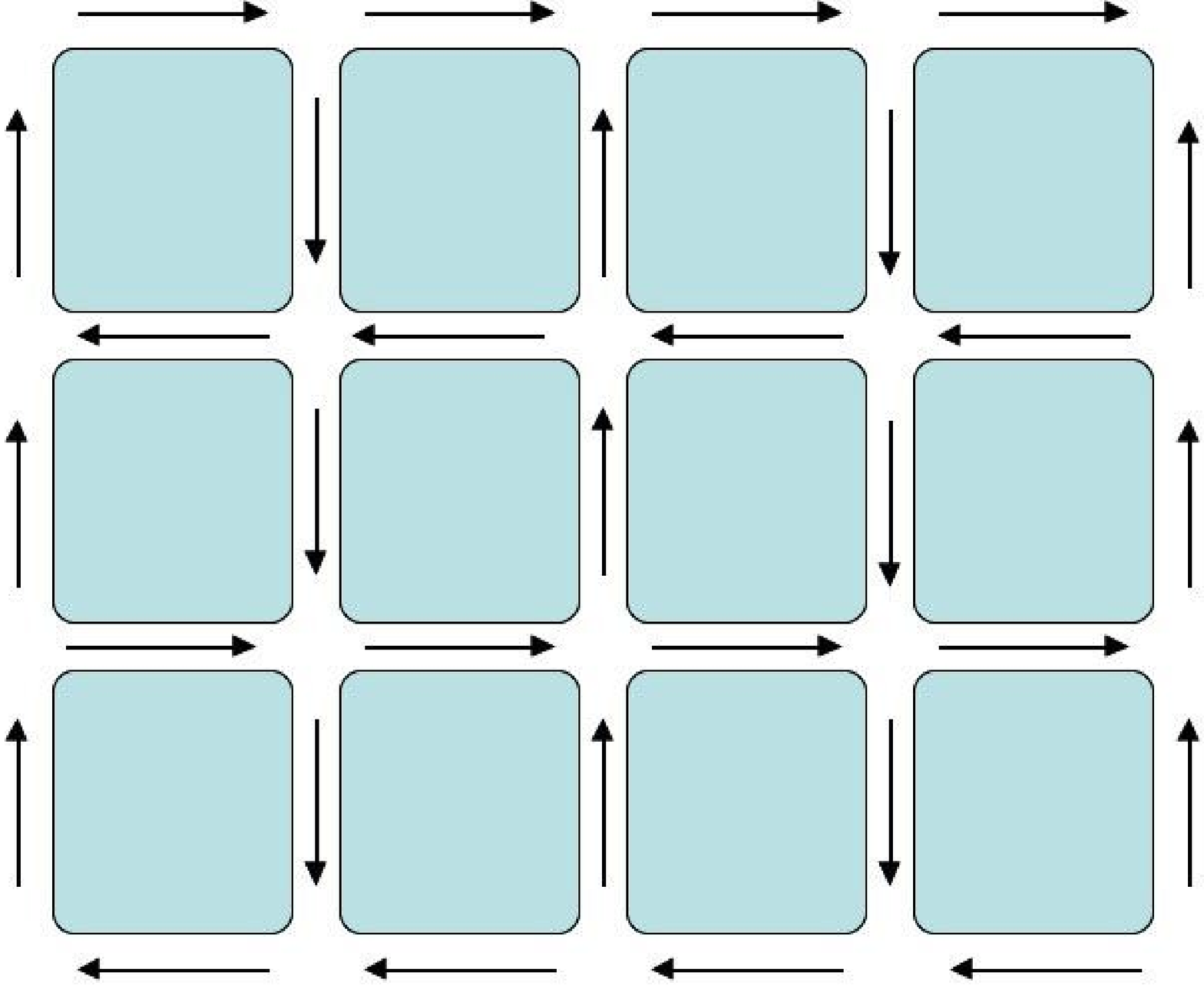}\\[2mm]
\includegraphics[width=6cm]{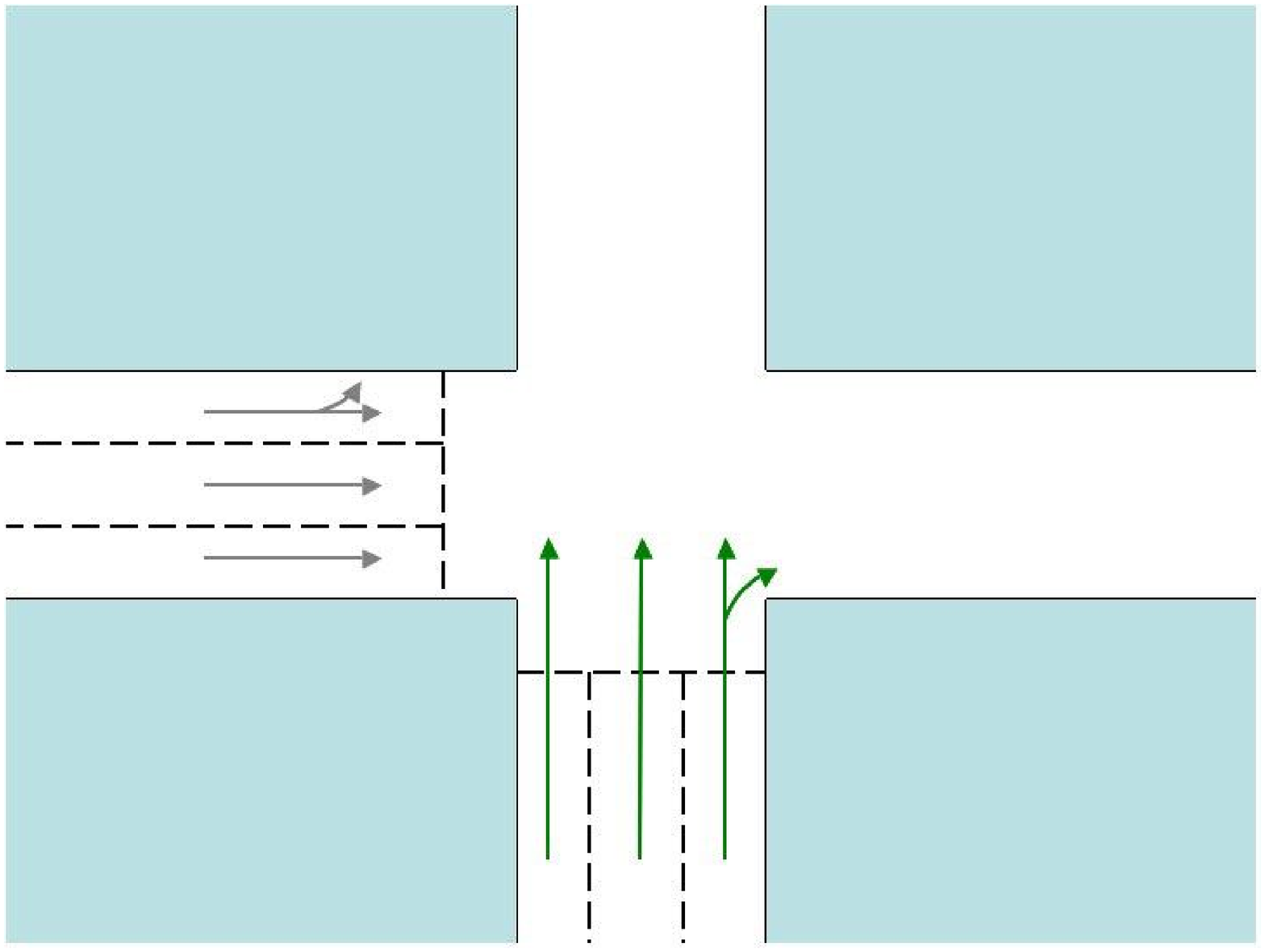}\\[2mm]
\includegraphics[width=6cm]{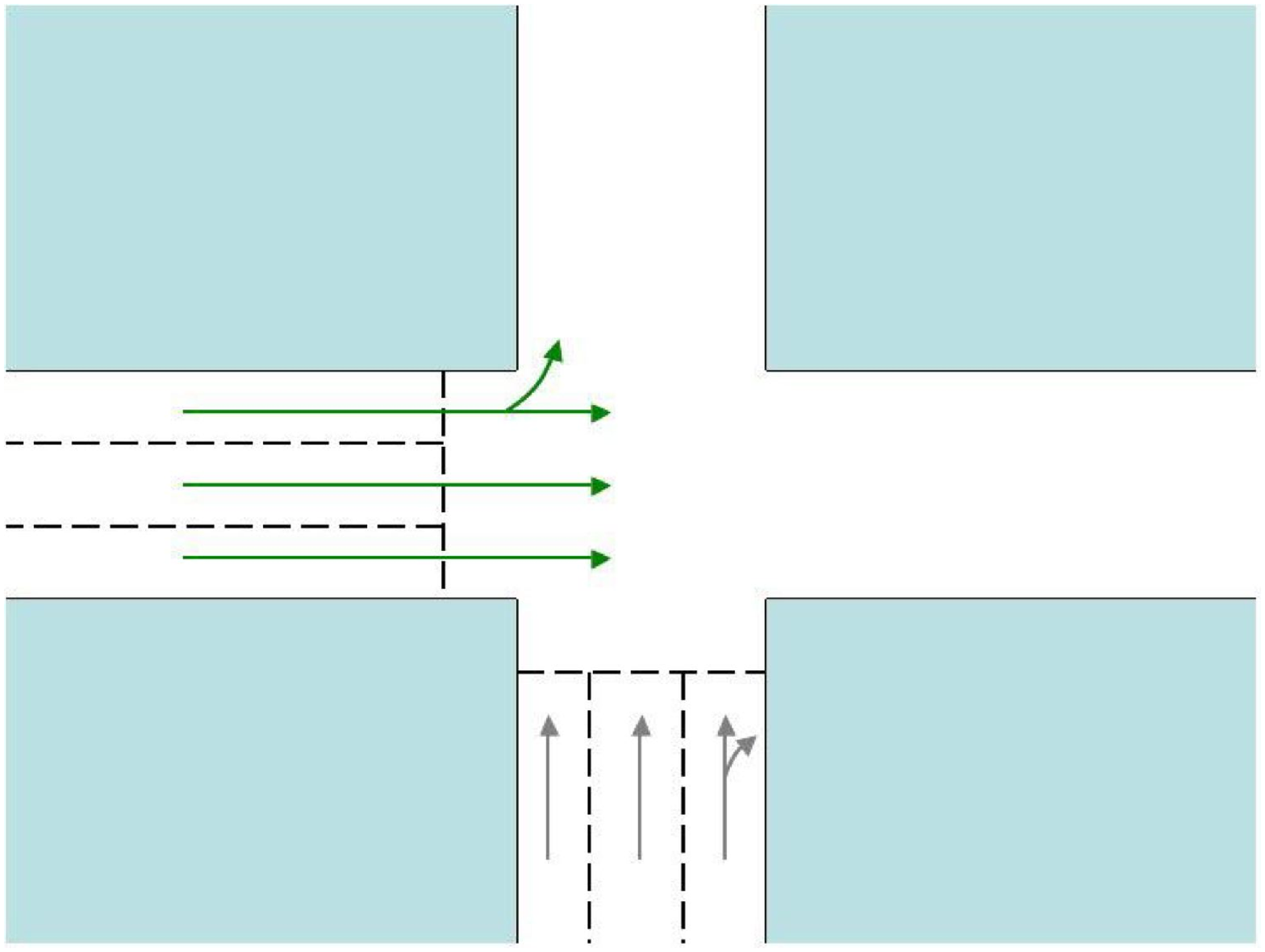}
\end{center}
\caption[]{Top: Schematic illustration of the unidirectional street layout in the center of Barcelona.
Center and bottom: Illustration of the two traffic phases, during which vehicles can move straight ahead or turn (either right or left, depending on the direction of the crossing road).}
\label{Illx}
\end{figure}

In this paper, we will first focus on the study of a single traffic intersection with uniform arrival flows, before we discuss later how to extent our control concept in various ways. Furthermore, for simplicity we will concentrate on the study of a traffic light control with {\it two} green phases only, which is generalized in Appendix \ref{Be}. As the traffic organization in parts of Barcelona shows, 
a two-phase control is {\it sufficient}, in principle, to reach all points in the road network: 
Just assume unidirectional flows in all streets with alternating
directions. Then, in each phase, traffic either flows straight ahead and/or turns (right or left,
depending on the driving direction in the crossing road). Hence, two intersecting unidirectional
roads imply two possible traffic phases, which alternate (see Fig. \ref{Illx}).
\par
While the optimization approach discussed in the following can be also applied to time-dependent arrival flows $A_1$ and $A_2$ per lane, when numerical solution methods are applied, for the sake of analytical tractability and closed formulas we will focus here on the case of constant flows over the short time periods involved in our optimization. $I_j$ will represent the number of lanes of road section $j$, and it will be assumed that vehicles passing a green light can freely enter the respective downstream road section. Like in Ref. \cite{Helbing2007,Helbing2003a,EPJB8}, the departure flows $\gamma_j(t)O_j(t)$ are specified as 
\begin{equation}
 \gamma_j(t)O_j(t) = \gamma_j(t) \left\{
 \begin{array}{ll}
 \widehat{Q}_j & \mbox{if } \Delta N_j(t) > 0, \\
 A_j(t-{\cal T}_j^0) &\mbox{otherwise.}
 \end{array}\right.
 \label{Oite}
\end{equation}
Herein, $\Delta N_j(t)$ denotes the number of delayed vehicles at time $t$ and $\widehat{Q}_j$ is the service rate per lane during the green phase. In the case of constant arrival rates $A_j$, the dependence on the time point $t$ and the shift by the free travel time ${\cal T}_j^0$ can be dropped. During amber and red time periods, the permeability $\gamma_j(t)$ is zero, as there is no outflow, while $\gamma_j(t)=1$ during green phases. Note that the departure flows $\gamma_j(t)O_j(t)$ may split up into a straigth and a turning flow after the traffic light, but for our further considerations, this is not relevant. 
\par
In the following, we will use some additional variables and parameters: $T_j$ shall denote the minimum green time, after which the vehicle queue in road section $j$ is fully dissolved (i.e. after which $\Delta N_j = 0$ and $O_j = A_j$).  In contrast, $\Delta T_j$ will stand for the actual green time period. Consequently, 
\begin{equation}
\Delta t_j = \Delta T_j - T_j 
\end{equation}
(if greater than zero) represents the excess green time, during which we have a free vehicle flow with $\gamma_j(t)O_j(t) = A_j$. $\tau_j$ shall be the switching time {\it before} the green phase $\Delta T_j$ for road section $j$. The sum 
\begin{equation}
T_{\rm cyc} = \tau_1 + \Delta T_1 + \tau_2 + \Delta T_2
\label{CyC}
\end{equation} 
is usually called the cycle time. Note, however, that we do not need to assume {\it periodic} operation. Within the framework of our model assumptions, we may consider {\it stepwise} constant flows. That is, the arrival flows may vary from one cycle (or even one green time period) to the next. Under such conditions, each green phase is adjusted to the changing traffic situation.

\section{Consideration of Traffic Flows}\label{flowopt}

The art of traffic control is to manipulate the permeabilities $\gamma_j(t)$ 
in a way that optimizes a given goal function. In fact, when the traffic volume is high enough, an oscillatory service corresponding to the operation of a traffic light can increase the effective intersection capacity as compared to the application of a first-come-first-serve rule for arriving vehicles \cite{Helbing2005,Helbing2007}: While the red and amber lights (corresponding to $\gamma_j(t) = 0$)
cause vehicles to queue up and wait, this implies a high flow rate and an
efficient service of vehicles when the traffic light turns green (i.e. $\gamma_j(t) = 1$). 
\par
One natural concept of traffic flow optimization would be to maximize the average overall throughput. 
This is measured by the function
\begin{equation}
 G_{\rm t}(t) = \frac{1}{t} \sum_j \int\limits_0^t dt' \; \gamma_j(t') O_j(t') \, . 
\end{equation}
Due to Eq. (\ref{Oite}), $G_{\rm t}(t)$ depends not only on the outflows $O_j(t)$, but also on the inflows $A_j(t)$ to the system. This makes $G_{\rm t}(t)$  basically dependent on the time-dependent 
origin-destination matrices of vehicle flows. 
\par
The numbers of vehicles accumulating during the red and amber time periods are
\begin{equation}
 I_1 \Delta N_1^{\rm max} = I_1 A_1 (\tau_2 + \Delta T_2 + \tau_1) 
\label{from1}
 \end{equation}
 and
 \begin{equation}
 I_2 \Delta N_2^{\rm max} = I_2 A_2 (\tau_1 + \Delta T_1 + \tau_2) \, , 
\end{equation}
where $\Delta N_j^{\rm max}$ represents the maximum number of delayed vehicles per lane in road section $j$, if the vehicle queue in it has been fully cleared before. $I_j$ is the number of lanes. As the service rate of queued vehicles during the green time $\Delta T_j$ is $\widehat{Q}_j$, and $A_j$ is the arrival rate of additional vehicles at the end of the queue, the mimimum green time required 
to dissolve the queue is given by 
\begin{equation}
 T_j = \frac{\Delta N_j^{\rm max}}{\widehat{Q}_j-A_j} \, .
\label{from2}
\end{equation}
From Eqs. (\ref{from1}) to (\ref{from2}) we obtain
\begin{equation}
 T_1 =  \frac{A_1}{\widehat{Q}_1 - A_1} \bigg(\tau_2 + \Delta T_2  + \tau_1 \bigg) \, .
 \end{equation}
Assuming $\Delta T_j = T_j$ (i.e. no excess green times) and inserting Eq. (\ref{from2}) yields
\begin{equation}
T_1 = \frac{A_1}{\widehat{Q}_1 - A_1} \bigg( \tau_2 + \frac{A_2(\tau_1 + T_1 + \tau_2)}{\widehat{Q}_2 - A_2} + \tau_1 \bigg)
 \end{equation}
for the clearing time $T_1$, or
 \begin{equation}
T_1 = (\tau_1 + \tau_2)  \frac{ \frac{A_1}{\widehat{Q}_1 - A_1} \left(1+ \frac{A_2}{\widehat{Q}_2 - A_2}\right)}
{1 - \frac{A_1A_2}{(\widehat{Q}_1-A_1)(\widehat{Q}_2-A_2)}} \, .
\end{equation}
With the analogous formula for $T_2$ we can determine the related cycle time, if
the traffic light turns red immediately when all queued vehicles have been served. After a
few intermediate mathematical steps, we finally get
\begin{equation}
 T^{\rm cyc} = \tau_1 + T_1 + \tau_2 + T_2
 = \frac{\tau_1+\tau_2}{1 - A_1/\widehat{Q}_1 - A_2/\widehat{Q}_2} \, .
\label{tcyc} 
\end{equation}
Moreover, one can show \cite{EPJB8}
\begin{equation}
 T_j = \frac{A_j }{\widehat{Q}_j} T^{\rm cyc} \, . \label{einfa}
\end{equation}
We can see that the cycle time and the clearing times $T_j$ diverge in the limit 
\begin{equation}
 \frac{A_1}{\widehat{Q}_1} + \frac{A_2}{\widehat{Q}_2} \rightarrow 1 \, .
 \label{capcond}
\end{equation}
If this expression (\ref{tcyc}) becomes negative, the vehicle queues in one or both ingoing road sections 
are growing larger and larger in time, as the intersection
does not have enough capacity to serve both arrival flows. See Ref. \cite{EPJB8} for a discussion of this case.
\par
Note that Eq. (\ref{tcyc}) determines the {\it smallest} cycle time that allows to serve all queued vehicles within the green time periods. Let us study now the effect of {\it extending} the green time periods $\Delta T_j$ beyond $T_j$:
The average throughput of the intersection is given by the overall flow of vehicles during one cycle time $T_{\rm cyc}=\tau_1+\Delta T_1+\tau_2 + \Delta T_2$. During that time period, a total number $(I_1 A_1+ I_2A_2)T_{\rm cyc}$ of vehicles is arriving in the two considered road sections. If all arriving vehicles are served during the cycle time $T_{\rm cyc}$, the average throughput is
\begin{equation}
 G_{\rm t} = \frac{(I_1 A_1+ I_2A_2) T_{\rm cyc}} 
 {T^{\rm cyc}} = I_1A_1 + I_2A_2 \, .
 \label{givenby}
\end{equation}
Therefore, in the case where we do not have an accumulation of vehicles over time,
which requires sufficient green times 
($\Delta T_j > T_j$) and a sufficient resulting service capacity
\begin{equation}
\frac{I_1\widehat{Q}_1\, \Delta T_1 + I_2\widehat{Q}_2 \, \Delta T_2}{T_{\rm cyc}}  \ge I_1A_1+I_2A_2 \, , 
\end{equation}
the throughput is determined by the sum $I_1A_1+I_2A_2$ of the overall arrival flows. Consequently, excess green times 
$ \Delta t_j = \Delta T_j - T_j > 0$ 
do not lead to smaller or larger intersection throughputs. But under what conditions should a green phase be extended, if at all? This shall be addressed in the next sections.

\section{Travel-Time-Oriented Signal Operation}\label{trati}

Rather than on a consideration of the flow, 
we will now focus on  the {\it cumulative waiting time}
\begin{equation}
 F(t) = \sum_j I_j  \int\limits_0^t  dt' \int\limits_0^{t'} dt^{\prime\prime}
 [A_j - \gamma_j(t^{\prime\prime})O_j(t^{\prime\prime})] 
\end{equation}
and minimize its average growth over a time period $t$ to be defined later. This corresponds to a minimization of the function
\begin{eqnarray}
 G(t) 
 &=& \frac{1}{t} \sum_j I_j  \int\limits_0^t  dt' \Delta N_j(t') \nonumber \\
&=& \frac{1}{t} \sum_j I_j  \int\limits_0^t  dt' \int\limits_0^{t'} dt^{\prime\prime}
 [A_j - \gamma_j(t^{\prime\prime})O_j(t^{\prime\prime})] \, , \quad
\label{instead}
\end{eqnarray}
which quantifies the time average of the overall delay time. The 
term on the right-hand side describes the increase of the overall waiting time
proportionally to the number $\Delta N_j$ of delayed cars, which is given by the
integral over the difference between the arrival and departure flows \cite{Helbing2007,EPJB8}. 

\subsection{The Optimize-One-Phase Approach}\label{OP}

When minimizing the goal function $G(t)$, it is essential upto what time $t$ we extend
the integral. In principle, it is possible to integrate over a full cycle or even many cycles of traffic operation, but the resulting formulas do not provide an intuitive understanding anymore. We will, therefore, focus on the optimization of a single phase, with full amber time periods $\tau_j$ in the beginning and $\tau_{j+1}$ at the end. This turns out to result in explicit and plausible formulas, while some other approaches we have tried, did not result in well interpretable results. Besides this practical aspect, when analytical results shall be obtained, the specification $t = \tau_1 + \Delta T_j + \tau_2$ chosen in the following makes sense: It ``charges'' the switching-related inefficiencies to the road that ``wants'' to be served. The switching of a traffic light should lead to a temporary {\it in}crease in traffic performance.
After completion of each green phase, the travel time optimization is repeated, so that one can compose the traffic light schedule as a sequence of optimized single phases (see Appendix  \ref{Be} for details).
\par
In Sec. \ref{MUL}, we will show that a multi-phase optimization yields better results, but requires a higher degree of sophistication. The treatment of situations with varying or pulsed traffic flows is even more difficult and can usually be solved only numerically. This issue is addressed in Ref. \cite{jstat}. 
\par
In our calculations, we will assume that the green time for road section 2 lasted for a time period $\Delta T_2$ and ended at time $t=0$. That is, we have now to determine the optimal duration $\Delta T_1$ of the green phase for road section 1 after an intermediate amber time period $\tau_1$. For this, we minimize the function
\begin{equation}
G_1(\tau_1 + \Delta T_1 + \tau_2) = 
\frac{F_1(\tau_1 + \Delta T_1 + \tau_2)} 
{\tau_1 + \Delta T_1 + \tau_2} \, , 
\end{equation} 
where the subscript ``1'' of $G$ and $F$ refers to road section 1, for which the green phase is determined. Assuming a step-wise constant outflow with $\gamma_jO_j= \widehat{Q}_j$, if $\Delta N_j > 0$, but $\gamma_jO_j = A_j$, if $\Delta N_j = 0$, and $\gamma_jO_j=0$, if $\gamma_j=0$, the integral over $t^{\prime\prime}$ results in a stepwise linear function, and the function $F_1(t)$ is characterized by quadradic dependencies. We will distinguish two cases: (a) The green time is potentially terminated {\it before} all queued vehicles have been served (i.e. $\Delta T_i \le T_i$), or (b) it is potentially {\it extended} (i.e. $\Delta T_i \ge T_i$). Let us start with the first case.
\begin{itemize}
\item[(a)] {\bf No excess green time} ($\Delta T_1 \le T_1$): In this case, $A_2(t^{\prime\prime}) - \gamma_2(t^{\prime\prime})O_2(t^{\prime\prime}) = A_2$ for $0 \le t^{\prime\prime} \le \tau_1 + \Delta T_1 + \tau_2$, i.e. over the period $\Delta T_1$ of the green time for road section 1 and the amber time periods $\tau_j$ and $\tau_{j+1}$ before and after it. In addition,
\begin{equation}
A_1 - \gamma_1(t^{\prime\prime})O_1(t^{\prime\prime}) = \left\{
\begin{array}{ll}
A_1 &  \mbox{if } 0\le t^{\prime\prime}<\tau_1, \\
A_1 - \widehat{Q}_1 & \mbox{if } \tau_1 \le t^{\prime\prime} < \tau_1 + \Delta T_1 , \\
A_1 & \mbox{otherwise.} 
\end{array}
\right.
\end{equation}
Using the abbreviation
\begin{equation}
 \Delta N_1^{\rm max} = \Delta N_1(\tau_1) = \Delta N_1(0) +  A_1 \tau_1  \, , 
 \label{abre}
\end{equation}
we get
\begin{eqnarray}
& &  F_1^{\rm a}(\tau_1 + \Delta T_1 + \tau_2) \nonumber \\
&=& I_1\bigg\{ \Delta N_1(0)\tau_1 + A_1 \frac{\tau_1{}^2}{2} \nonumber \\
& & + \Delta N_1^{\rm max} \Delta T_1 - (\widehat{Q}_1-A_1)\frac{\Delta T_1{}^2}{2} \nonumber \\
& & + [\Delta N_1^{\rm max} - (\widehat{Q}_1-A_1)\Delta T_1]\tau_2 + A_1 \frac{\tau_2{}^2}{2} \bigg\} \nonumber \\
&+& I_2 \! \left[ \Delta N_2(0) (\tau_1 + \!\Delta T_1 + \tau_2) + \!A_2 \frac{(\tau_1 + \!\Delta T_1+\tau_2)^2}{2} \right] \nonumber \\
&=& I_1 \bigg[ \Delta N_1(0) (\tau_1+\Delta T_1 + \tau_2) + \frac{A_1}{2} (\tau_1 + \Delta T_1+\tau_2)^2 \nonumber \\
&& - \frac{\widehat{Q}_1}{2} \Delta T_1(\Delta T_1 + 2\tau_2) \bigg] \nonumber \\
&+& I_2 \! \left[ \Delta N_2(0) (\tau_1 + \!\Delta T_1 + \tau_2) + \!A_2 \frac{(\tau_1 + \!\Delta T_1+\tau_2)^2}{2} \right] ,
\nonumber \\ & & 
\end{eqnarray}
where the superscript ``a'' refers to case (a).
Dividing the above function by $(\tau_1 + \Delta T_1 + \tau_2)$ and making the plausible assumption $\tau_1 = \tau_2$ of equal amber time periods for simplicity, we gain 
\begin{eqnarray}
G_1^{\rm a}(\tau_1 + \Delta T_1 + \tau_2) 
&=& I_1 \bigg[ \Delta N_1(0) + \widehat{Q}_1 \tau_2 \nonumber \\
 & & - (\widehat{Q}_1-A_1) \frac{\tau_1 + \Delta T_1+\tau_2}{2} \bigg] \nonumber \\
&+& I_2 \!\left[ \Delta N_2(0) + \!A_2 \frac{\tau_1 + \Delta T_1+ \tau_2}{2} \right] . \nonumber \\
& & \label{GAIN1}
\end{eqnarray}
If $I_1(\widehat{Q}_1-A_1) < I_2A_2$, i.e. when the number of queued vehicles in road section 2 grows faster than it can be reduced in road section 1, the minimum of this function is reached for $\Delta T_1 = 0$, corresponding
to a situation where it is not favorable to turn green for section $j=1$. For 
\begin{equation}
I_1(\widehat{Q}_1-A_1) > I_2A_2\, , 
\label{howdi1}
\end{equation}
the value of $G_1^{\rm a}$ goes down with
growing values of $\Delta T_1$, and the minimum is reached for a value $\Delta T_1 \ge T_1$.
\par
\item[(b)] {\bf  Potential green time extension} ($\Delta T_1 \ge T_1$): Let us assume that we (possibly) have an excess green time, i.e. $\Delta t_i = \Delta T_i -T_i \ge 0$. In this case,
\begin{equation}
A_1 - \gamma_1(t^{\prime\prime})O_1(t^{\prime\prime}) = \left\{
\begin{array}{ll}
A_1 &  \mbox{if } 0\le t^{\prime\prime}<\tau_1, \\
A_1 - \widehat{Q}_1 & \mbox{if } \tau_1 \le t^{\prime\prime} < \tau_1 + T_1 , \\
A_1 &  \mbox{if } t^{\prime\prime} \ge  \tau_1 + \Delta T_1 , \\ 
0 & \mbox{otherwise.}
\end{array}
\right.
\end{equation}
Considering that now, $\Delta N_1(t') = 0$ for $\tau_1 + T_1 \le t' < \tau_1 + \Delta T_1$, and introducing the {\it clearing time}
\begin{equation}
 T_1 = \frac{\Delta N_1^{\rm max}}{\widehat{Q}_1 - A_1} =  \frac{\Delta N_1(0) + A_1\tau_1}{\widehat{Q}_1 - A_1} \, ,  
 \label{consu}
\end{equation}
we obtain 
\begin{eqnarray}
& & F_1^{\rm b}(\tau_1 + \Delta T_1 + \tau_2) \nonumber \\
 &=& I_1 \bigg[ \Delta N_1(0) \tau_1 + A_1 \frac{\tau_1{}^2}{2} 
+ \Delta N_1^{\rm max} T_1 \nonumber \\
& & - (\widehat{Q}_1-A_1)\frac{T_1{}^2}{2} + A_1 \frac{\tau_2{}^2}{2} \bigg] \nonumber \\
&+& I_2\left[ \Delta N_2(0) (\tau_1+ \Delta T_1 + \tau_2) + A_2 \frac{(\tau_1+ \Delta T_1+\tau_2)^2}{2} \right] \nonumber \\
 &=& I_1 \left[ \Delta N_1^{\rm max} \tau_1 + \frac{A_1}{2}(\tau_2{}^2 - \tau_1{}^2)+ \frac{(\Delta N_1^{\rm max})^2}{2(\widehat{Q}_1 - A_1)} \right] \nonumber \\
 &+& I_2\left[ \Delta N_2(0) (\tau_1+ \Delta T_1 + \tau_2) + A_2 \frac{(\tau_1+\Delta T_1+\tau_2)^2}{2} \right]  
 \nonumber \\ & &  
\label{equus}
\end{eqnarray}
Assuming again $\tau_1= \tau_2$ for simplicity, introducing the abbreviation
\begin{equation}
 E_1 = \Delta N_1^{\rm max} \tau_1 + \frac{(\Delta N_1^{\rm max})^2}{2(\widehat{Q}_1 - A_1)} \, , 
 \label{E1}
\end{equation} 
and dividing Eq. (\ref{equus}) by $(\tau_1+ \Delta T_1 + \tau_2)$ yields
\begin{eqnarray}
 & & G_1^{\rm b}(\tau_1+\Delta T_1 + \tau_2) \nonumber \\
 &=& \frac{I_1E_1}{\tau_1+\Delta T_1 + \tau_2} \nonumber \\ 
&+& I_2\left[ \Delta N_2(0) + A_2 \frac{\tau_1 + \Delta T_1 + \tau_2}{2} \right] \, .\qquad 
\label{GAIN2}
\end{eqnarray}
This expression shall be minimized under the constraint $\Delta T_1 \ge T_1$. In order to determine the minimum,
we set the derivative with respect to $\Delta T_1$ to zero and get
\begin{equation}
 0 = \frac{d G_1^{\rm b}(\tau_1+ \Delta T_1 + \tau_2)}{d\,\Delta T_1}
 = - \frac{I_1E_1}{(\tau_1 + \Delta T_1+\tau_2)^2} + \frac{I_2A_2}{2} \, .
\end{equation}
The minimum is located at
\begin{equation}
 (\tau_1+ \Delta T_1+ \tau_2)^2 = \frac{2I_1E_1}{I_2A_2} \, ,
\label{wit}
\end{equation}
if $\Delta T_1 \ge T_1$. Considering Eq. (\ref{consu}), $\Delta T_1 \ge T_1$ implies 
\begin{equation}
 (\tau_1 + \Delta T_1+\tau_2)^2 \ge 
 \left( \tau_1+ \frac{\Delta N_1^{\rm max}}{\widehat{Q}_1-A_1} + \tau_2\right)^2 \, .
\label{impli}
\end{equation}
With Eq. (\ref{wit}) this leads to the condition
\begin{eqnarray}
 & & \frac{(\Delta N_1^{\rm max})^2}{\widehat{Q}_1 - A_1}  \left( \frac{I_1}{I_2A_2} - \frac{1}{\widehat{Q}_1 - A_1}\right) \nonumber \\
 & & + 2 \Delta N_1^{\rm max} \left( \frac{I_1\tau_1}{I_2 A_2} - \frac{\tau_1+\tau_2}{\widehat{Q}_1-A_1}\right) 
 \ge (\tau_1+\tau_2)^2 \, .\quad
 \label{verysmall}
\end{eqnarray}
If inequality (\ref{verysmall}) is not fulfilled, we must have $\Delta T_1 < T_1$.
\end{itemize}
For completeness, we note that 
\begin{equation}
G_1^{\rm a} (\tau_1 + T_1 +\tau_2) = G_1^{\rm b}(\tau_1 + T_1 + \tau_2) \, , 
\end{equation}
i.e. the goal function $G_1$ is continuous in $\Delta T_1 = T_1$, while it must not be smooth.
Moreover, $\Delta N_1(0) = A_1(\tau_2+\Delta T_2)$ and $\Delta N_2(0) = 0$, if the vehicle queues have been fully cleared before the traffic light is switched. The case where the queue is not fully dissolved is treated in Ref. \cite{EPJB8}.

\subsection{Transformation to Dimensionless Variables and Parameters}\label{sca}

For an analysis of the system behavior, it is useful to transform variables and parameters to dimensionless units. Such dimensionless units are, for example, the capacity utilizations
\begin{equation}
u_i = \frac{A_i}{\widehat{Q}_i} \label{udef}
\end{equation}
of the road sections $i$ and the relative size 
\begin{equation}
\kappa = \frac{I_1A_1}{I_2A_2} = \frac{I_1u_1\widehat{Q}_1}{I_2u_2\widehat{Q}_2}
= \frac{u_1}{u_2} K
\end{equation}
of the arrival flows, where 
\begin{equation}
K = \frac{I_1\widehat{Q}_1}{I_2\widehat{Q}_2} \, . \label{Ka}
\end{equation}
Furthermore, we may scale the green times $\Delta T_i$ by the sum of amber time periods $\tau_1+\tau_2$, which defines the dimensionless green times
\begin{equation}
\sigma_i = \frac{\Delta T_i}{\tau_1+\tau_2} 
\end{equation} 
and the dimensionless clearing times
\begin{equation}
\hat{\sigma}_j =  \frac{T_j}{\tau_1+\tau_2} 
= \frac{\Delta N_1^{\rm max}}{(1-u_1)\widehat{Q}_i(\tau_1+\tau_2)} \, . \label{diclear}
\end{equation}
In order to express the previous relationships exclusively by these quantities, we must consider
that a number $A_1(\tau_2 + \Delta T_2 +\tau_1)$ of vehicles per lane accumulates during the time period $(\tau_2 + \Delta T_2 +\tau_1)$, in which the vehicle flow on road section 1 is not served.
With Eq. (\ref{abre}) this implies 
\begin{equation}
\Delta N_1^{\rm max} = \Delta N_1(0) + A_1\tau_1= A_1 (\tau_2 + \Delta T_2 + \tau_1) \, , 
\label{assu}
\end{equation}
if the vehicle queue in road section 1 has been fully cleared during the previous green time. Then, we have
\begin{equation}
 \frac{\Delta N_1^{\rm max}}{\tau_1+\tau_2} = A_1 (1 + \sigma_2) \, , 
\end{equation}
and from Eqs. (\ref{E1}) and (\ref{GAIN1}) we get
\begin{equation}
 \frac{2E_1}{(\tau_1+\tau_2)^2} = A_1 (1+\sigma_2)\frac{2\tau_1}{\tau_1+\tau_2}
 + \frac{(A_1)^2 (1+\sigma_2)^2}{\widehat{Q}_1 - A_1} \, .  
\end{equation}
\pagebreak
With $A_1 = u_1 \widehat{Q}_1$ and $\tau_1 = \tau_2$, Eq. (\ref{wit}) belonging to the case of extended green time for road section 1 can be written as 
\begin{equation}
(1+\sigma_1)^2 = [1+\tilde{\sigma}_1(\sigma_2)]^2 =\kappa \bigg[ (1+\sigma_2) + \frac{u_1}{1-u_1}(1+\sigma_2)^2\bigg] \, .
\label{RES2}
\end{equation}
The solution of this equation defines the relationship $\tilde{\sigma}_1(\sigma_2)$ for the optimal scaled green time period $\sigma_1$ as a function of $\sigma_2$, if the green time for road section 1 is extended. Moreover, in dimensionless variables, the condition (\ref{verysmall}) for green time extension becomes 
\begin{equation}
 \frac{u_1}{1-u_1}(1+\sigma_2)^2 \left( \kappa - \frac{u_1}{1-u_1} \right) + (1+\sigma_2) \left( \kappa - \frac{2u_1}{1-u_1}\right)  \ge  1
\end{equation}
or
\begin{equation}
 \left[ \frac{u_1(1+\sigma_2)^2}{1-u_1} + (1+\sigma_2)\right] \!\!\left( \kappa - \frac{u_1}{1-u_1} \right) \ge 
 \frac{1+u_1\sigma_2}{1-u_1}  \, . \label{COND2}
\end{equation}
However, we can check for green time extension also in a different way, since the extension condition $\Delta T_1 > T_1$ can be written as $\tilde{\sigma}_1 > \hat{\sigma}_1$. Using
Eqs. (\ref{consu}), (\ref{diclear}) and (\ref{assu}), the dimensionless green time $\sigma_1$ for the case of no green time extension may be presented as
\begin{equation}
\sigma_1 = \hat{\sigma}_1(\sigma_2) = \frac{A_1(1+\sigma_2)}{\widehat{Q}_1 - A_1}
= \frac{u_1(1+\sigma_2)}{1-u_1} 
\label{RES1}
\end{equation}
or
\begin{equation}
1 + \sigma_1 = \frac{1+u_1\sigma_2}{1-u_1} \, . \label{RES1a}
\end{equation}
Moreover, from $\sigma_1 = \hat{\sigma}_1(\sigma_2)$ follows 
\begin{equation}
\frac{\sigma_1}{1+\sigma_1+\sigma_2} = \frac{u_1(1+\sigma_2)}{(1-u_1)\Big[ 1 + \frac{u_1}{1-u_1}(1+\sigma_2) + \sigma_2\Big]} = u_1 \, . \label{uu1}
\end{equation}
That is, in the case where road section 1 is completely cleared, but there is no green time extension, the green time fraction 
\begin{equation}
\frac{\Delta T_1}{T_{\rm cyc}} = \frac{\sigma_1}{1+\sigma_1+\sigma_2}
\label{grtifrac}
\end{equation}
agrees with the utilization $u_1$. Moreover, one can show
\begin{equation}
\frac{\partial}{\partial \sigma_1} \left( \frac{\sigma_1}{1+\sigma_1+\sigma_2}\right) 
= \frac{1+\sigma_2}{(1+\sigma_1+\sigma_2)^2} > 0 \, . 
\end{equation}
Therefore, $\sigma_1 > \hat{\sigma}_1$ implies a green time fraction greater than $u_1$, and we have excess green time for road section~1, if
\begin{equation}
\frac{\tilde{\sigma}_1(\sigma_2)}{1+\tilde{\sigma}_1(\sigma_2) + \sigma_2} > u_1 \, . \label{COND3}
\end{equation}
An analogous condition must be fulfilled, if excess green times on road section 2 shall be optimal. It reads
\begin{equation}
\frac{\tilde{\sigma}_2(\sigma_1)}{1+ \sigma_1+\tilde{\sigma}_2(\sigma_1) } > u_2 \, , \label{COND3a}
\end{equation}
where
\begin{equation}
[1+\tilde{\sigma}_2(\sigma_1)]^2 =\frac{1}{\kappa} \bigg[ (1+\sigma_1) + \frac{u_2}{1-u_2}(1+\sigma_1)^2\bigg] \, , \label{COND5}
\end{equation}
which has been gained by interchanging indices 1 and 2 and replacing $\kappa$ by $1/\kappa$ in Eq. (\ref{RES2}).

\subsection{Control Strategies and Slower-is-Faster Effect} \label{SF2}

\begin{figure*}[htbp]
\begin{center}
\hbox{\includegraphics[width=9.2cm]{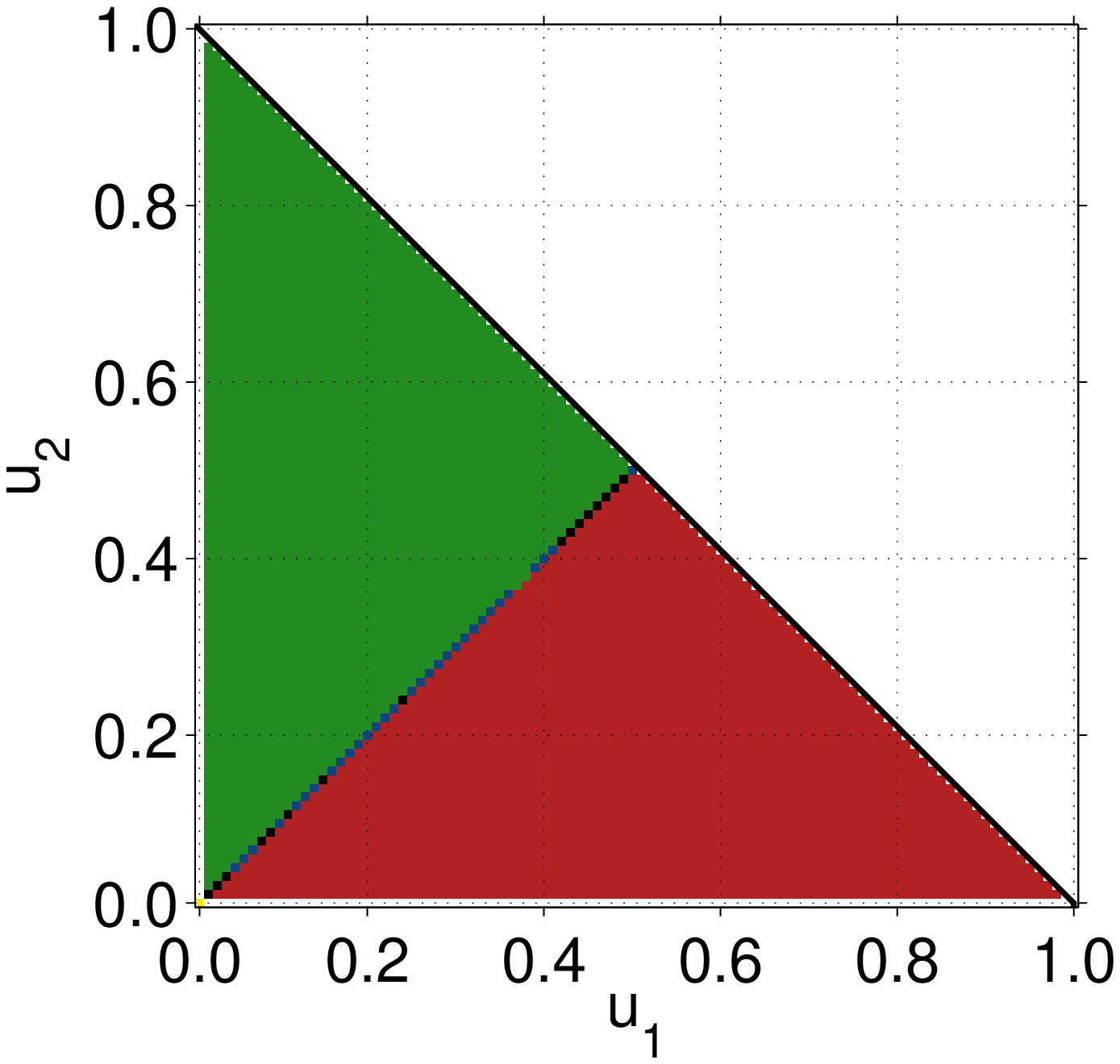}
\includegraphics[width=9.2cm]{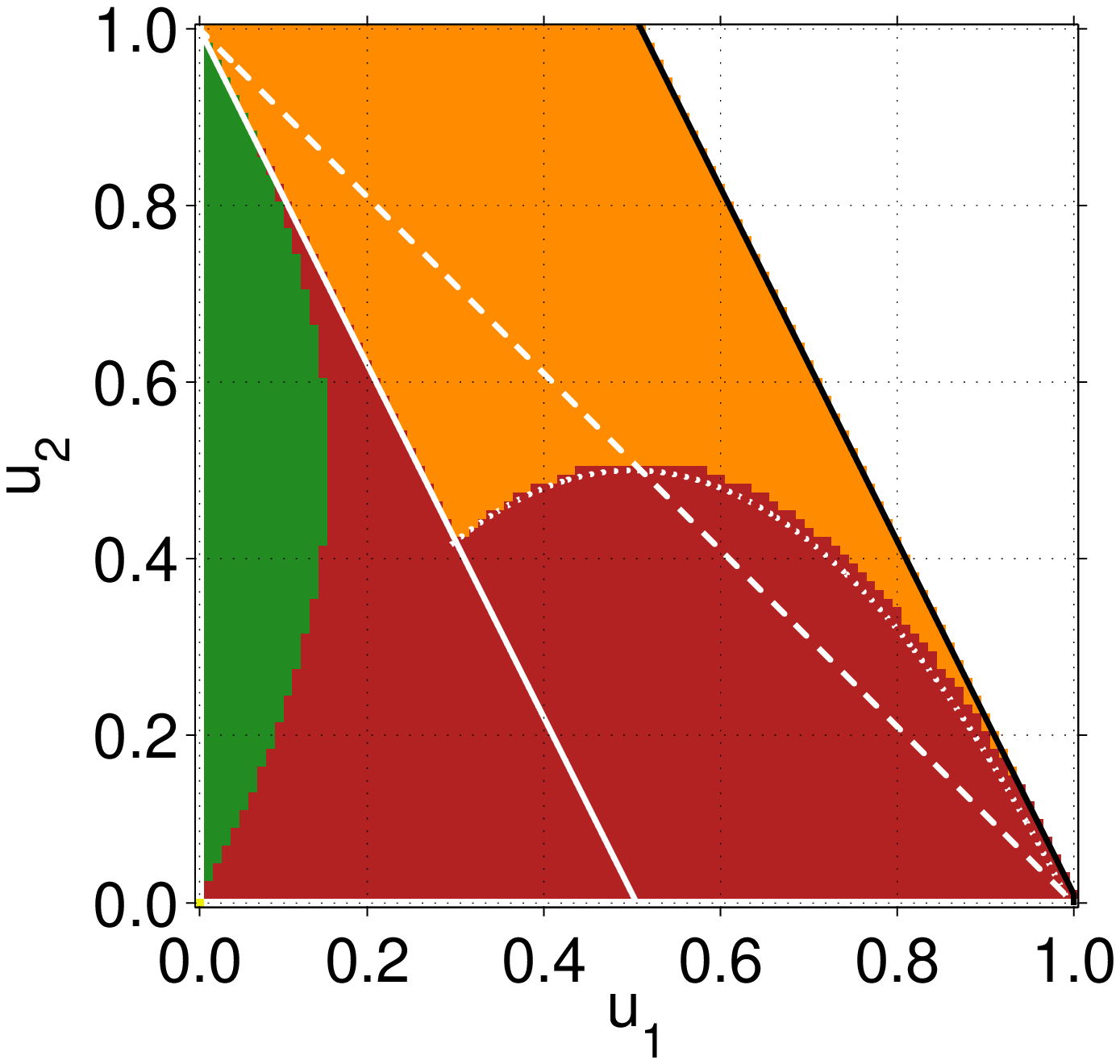}}
\end{center}
\caption[]{Operation regimes of (periodic) signal control for $K=1$ (left) and $K=3$ (right) as a function of the utilizations $u_j$ of both roads $j$ according to the one-phase travel time optimization approach. For each combination of $u_1$ and $u_2$, the operation regime has been determined after convergence of the signal control procedure described in Appendix \ref{Be}. The separating lines are in good agreement with our analytical calculations. For example, the solid falling lines are given by Eq. (\ref{COND4a}), while the dotted parabolic line in the right illustration corresponds to $u_2 = K u_1(1-u_1)$ and results by equalizing Eq. (\ref{RES2}) with the square of Eq. (\ref{RES1a}), assuming $\sigma_2 = 0$ (i.e. no service of road section 2).
The different operation regimes are characterized as follows: In the green triangular or parabolic area to the left of both illustrations, where the utilization $u_1$ of road section 1 is sufficiently small, the service of road section 2 is extended. In the adjacent red area below the white area (left) or the solid line (right), road section 2 is just cleared, while above the separating line $u_2 = 1 - Ku_1$, road section 2 is not served at all. Road section 1, in constrast, gets just enough green time to clear the vehicle queue in the green area (and the orange area towards the top of the right illustration), while it gets extended green time in the red area towards the bottom, where the utilization $u_2$ of road section 2 is sufficiently small. In the white area given by $u_2 > K(1-u_1)$, road section 1 gets no green time anymore.  Between the dashed and the solid white lines, road section 2 is not served, although there would be enough capacity to satisfy the vehicle flows in both roads. Improved operation regimes are presented in Fig. \ref{displ5}.} 
\label{newfig}
\end{figure*}
Based on the results of Sec. \ref{OP} and the scaled formulas of Sec. \ref{sca}, we can 
now formulate control strategies for a single traffic light within the optimize-one-phase approach: 
\begin{itemize}
\item[(i)] {\bf Terminate the green light for road section 1 immediately}, corresponding to $\sigma_1 = 0$, if condition (\ref{howdi1}) is violated, i.e. if
\begin{equation}
1 - u_1 \le  \frac{u_2}{K}  
 \label{COND4} 
\end{equation}
is fulfilled.
To obtain the dimensionless form of this inequality, we have considered $A_j = u_j I_j\widehat{Q}_j$ and Eq. (\ref{Ka}). In case (i),  travel time optimization for one phase advises against turning green for road section 1. Of course, in reality, drivers cannot be stopped forever. Either, one would have to give them a short green phase after a maximum tolerable time period, or at least one would have to allow vehicles to turn on red, i.e. to merge the crossing flow, whenever there is a large enough gap between two successive vehicles. Alternatively, one may apply an optimize-multiple-phases approach, see Sec.~\ref{MUL}. It implies a service of side roads even when the intersection capacity is insufficient to satisfy all inflows completely.
\item[(ii)] {\bf Terminate the green phase for road section 1, when the vehicle queue is completely resolved}, if conditions (\ref{COND4}) and (\ref{COND2}) are violated. In this case, the scaled green time $\sigma_1$ is given by Eq. (\ref{RES1}).
\item[(iii)] {\bf Extend the green times for road section 1} in accordance with formula (\ref{RES2}), if
the condition (\ref{COND2}) is fulfilled. The recommended delay in the switching time constitutes a {\it slower-is-faster effect}. In this situation, it takes some additional time to accumulate enough vehicles on road section 2 to guarantee an efficient service in view of the inefficiencies caused by the switching times $\tau_j$.
\end{itemize}
In Fig. \ref{newfig}, operation regime (i) is indicated in white and operation regime (iii) in red,  while
operation regime (ii) is shown in green, if road section 2 is served, otherwise in orange.

\subsection{Operation Regimes for Periodic Operation} \label{peropreg}

In the previous section, we have determined the optimal green time period $\sigma_1$ for road section 1, assuming that the last green time period $\sigma_2$ for road section 2 and $N_1(0)$ were given. Of course, $\sigma_1$ will then determine $\sigma_2$, etc. If the utilizations $u_j$ are constant and not too high, the sequence of green phases converges towards a {\it periodic} signal operation (see Fig. \ref{converge}). It will be studied in the following. While the formulas for the determination of $\sigma_1$ were derived in Sec. \ref{sca}, the corresponding formulas for $\sigma_2$ can be obtained by interchanging the indices 1 and 2 and replacing $\kappa$ by $1/\kappa$ in all formulas. 
In principle, there could be the following cases, if we restrict ourselves to reasonable solutions with $\sigma_j \ge 0$:
\begin{figure}[htbp]
\centering
\includegraphics[width=9.2cm]{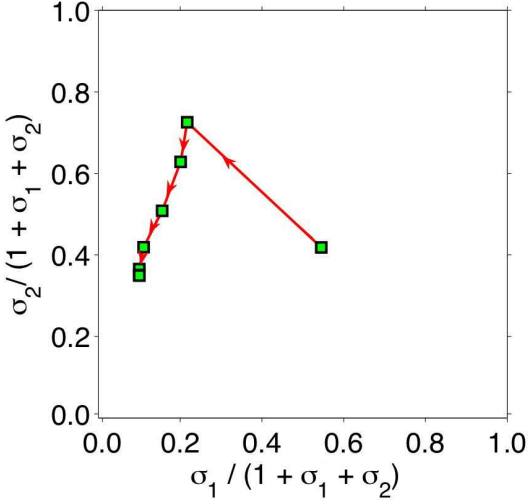}
\caption[]{Green time fraction $\sigma_2/(1+ \sigma_1+\sigma_2)$ for road section 2 vs. green time fraction $\sigma_1/(1+ \sigma_1+\sigma_2)$ for road section 1, if we apply the signal control algorithm described in Appendix \ref{Be} to a randomly chosen initial queue length $\Delta N_1(0)$ in road section 1 and $K=2$ (i.e. road section 1 has 2 times as many lanes as road section 2). One can clearly see that the green time fractions quickly converge towards values that do not change anymore over time. The solution corresponds to periodic signal operation.}
\label{converge}
\end{figure}
\begin{itemize}
\item[(0)] According to travel time minimization, one or both road sections should not be served, 
if (\ref{COND4}) is fulfilled for one or both of the road sections. This case occurs if
\begin{equation}
1 - u_1 - \frac{u_2}{K}  \le 0 \quad \mbox{or} \quad 1 - Ku_1 - u_2 \le 0 
\label{COND4a}
\end{equation}
(see the area above the white solid line in the right illustration of Fig. \ref{newfig}). According to this, service should focus on the main flow, while crossing flows should be suppressed, thereby enforcing a re-routing of traffic streams when this would be favorable to minimize travel times. Of course, in such situations vehicles should still be allowed to turn on red and to merge the crossing flow, when vehicle gaps are large enough.
\item[(1)] Both green time periods are terminated as soon as the respective vehicle queues are fully dissolved. In this case, we should have the relationships $\sigma_1 = \hat{\sigma}_1(\sigma_2)$ and 
$\sigma_2 = \hat{\sigma}_2(\sigma_1)$, where $\hat{\sigma}_j$ is defined in Eq. (\ref{RES1}). 
After a few steps, the condition $\sigma_1 =\hat{\sigma}_1(\hat{\sigma}_2(\sigma_1))$ implies
\begin{equation}
\sigma_j = \hat{\sigma}_j = \frac{u_j}{1-u_1-u_2} 
\end{equation}
and
\begin{equation}
\frac{\sigma_j}{1+\sigma_1+\sigma_2} = u_j \, . \label{gtf}
\end{equation}
According to Eq. (\ref{gtf}), the green time fraction of each road section in case (1) should be proportional to the respective utilization $u_j$ of the flow capacity.
\item[(2)] Road section 2 gets an excess green time, while the green phase of road section 1 ends after the dissolution of the vehicle queue (see green area in Fig. \ref{newfig}). In this case we should have $\sigma_1=\hat{\sigma}_1(\tilde{\sigma}_2(\sigma_1))$, where $(1+\tilde{\sigma}_2)$ is defined by formula (\ref{COND5}).
This gives
\begin{equation}
\sigma_1 = \frac{u_1}{1-u_1} \sqrt{ \frac{1}{\kappa} \Big( (1+\sigma_1) + \frac{u_2}{1-u_2}(1+\sigma_1)^2\Big)} \, ,
\end{equation}
which eventually leads to a quadratic equation for $\sigma_1$, namely
\begin{eqnarray}
& & \big[u_1u_2{}^2 - K(1-u_1)^2(1-u_2)\big]\sigma_1{}^2 \nonumber \\
& & + u_1 u_2 (1+u_2)\sigma_1 + u_1u_2 = 0 \, . 
\label{ACCORD}
\end{eqnarray}
To determine $\sigma_2$, we can either use the relationship $\sigma_2 = \tilde{\sigma}_2(\sigma_1)$ or invert the formula $\sigma_1 = \hat{\sigma}_1(\sigma_2)$. Doing the latter, Eq. (\ref{RES1}) gives
\begin{equation}
\sigma_2 = \frac{1-u_1}{u_1} \sigma_1 - 1 \, . 
\end{equation}
According to Eqs. (\ref{uu1}) and (\ref{COND3a}), the occurence of case (2) requires that the resulting solution satisfies
\begin{equation}
\frac{\sigma_1}{1+\sigma_1+\sigma_2} = u_1 \quad \mbox{and} \quad
\frac{\sigma_2}{1+\sigma_1+\sigma_2} > u_2 \, . \label{atisf}
\end{equation}
\item[(3)] Road section 1 gets an excess green time, while the green phase of road section 2 ends after the dissolution of the vehicle queue (see red area in Fig. \ref{newfig}). The formulas for this case are obtained from the ones of case (2) by interchanging the indices 1 and 2 and replacing $\kappa$ by $1/\kappa$.
\item[(4)] Both road sections get excess green time periods. This case would correspond to $\sigma_1=\tilde{\sigma}_1(\tilde{\sigma}_2(\sigma_1))$, and the solutions should fulfil
\begin{equation}
\frac{\sigma_1}{1+\sigma_1+\sigma_2} > u_1 \quad \mbox{and} \quad
\frac{\sigma_2}{1+\sigma_1+\sigma_2} > u_2 \, . 
\label{EXI}
\end{equation}
\end{itemize}
According to numerical results (see Fig. \ref{newfig}), cases (0), (2), and (3) do all exist, while the conditions for cases (1) and (4) are not fulfilled. Note, however, that small vehicle flows should better be treated as discrete or pulsed rather than continuous flows, in order to reflect the arrival of single vehicles (see Ref. \cite{TGF05} for their possible treatment within a continuous flow framework). In other words, for rare vehicle arrivals, we either have $u_1 > 0$ and $u_2 = 0$, or we have $u_2 > 0$ and $u_1 = 0$. Hence, the case of small utilizations $u_j$ will effectively imply green time extensions for both road sections due to the discreteness of the flow, and it allows single vehicles to pass the traffic light without previously stopping at the red light.
\par
Summarizing the above, one-phase optimization provides extra green times for road sections, as long as both of them are fully served. While in one road section, this slower-is-faster effect allows some vehicles to pass the traffic light without stopping, in the other road section it causes the formation of a longer vehicle queue, which supports an efficient service of a substantial number of vehicles after the traffic light turns green. In this connection, it is useful to remember that switching is costly due to the amber times, which are ``lost'' service times.

\subsection{Minimization of Vehicle Queues}\label{opq}

We have seen that travel time minimization implies the possibility of case (0), where one of the road sections (the side road) in not being served. This case should not occur as long as the
intersection capacity is not fully used. According to Eqs.  (\ref{capcond}) and (\ref{udef}), the intersection capacity is sufficient, if 
\begin{equation}
u_1+u_2 \le 1 
\label{DIFF}
\end{equation}
As the inequalities (\ref{COND4a}) and (\ref{DIFF}) do not agree, conditions may occur, where the vehicle queue in one road section (a side road) continuously increases, even though the intersection capacity would allow to serve both flows (see the orange and red areas above the dashed white line in Fig. \ref{newfig}). This can result in an ``unstable'' signal control scheme, which causes undesired spillover effects and calls for a suitable stabilization strategy \cite{jstat}. As we will see in the following, this problem can be overcome by minimizing vehicle queues rather than travel times.
\par
Conditions (\ref{COND4a}) and (\ref{DIFF}) agree, if $K=1$, particularly when $I_1 = I_2$ and $\widehat{Q}_1 = \widehat{Q}_2$. Therefore, let us assume this case in the following, corresponding to
\begin{equation}
\kappa = \frac{I_1A_1}{I_2A_2} = \frac{I_1u_1\widehat{Q}_1}{I_2u_2\widehat{Q}_2} = \frac{u_1}{u_2} \, . 
\end{equation} 
$\widehat{Q}_1 = \widehat{Q}_2$ holds, when the street sections downstream of the intersection do not impose a bottleneck. Furthermore, $I_1 = I_2 = 1$ corresponds to a minimization of the average {\it queue length} rather than the average delay time. Such a minimization of the queue length makes a lot of sense and means that the optimization is made from the perspective of the traffic network rather than from the perspective of the driver. This minimizes spillover effects and, at the same time, keeps travel times low.  

\subsection{Complexity of Traffic Light Control}

It is interesting that already a single intersection with constant arrival flows shows a large variety of operation regimes. In order to get an idea of the complexity of optimal traffic light control in general,
let us ask about the dimension of the phase space. For such an analysis, it is 
common to transform all parameters to dimensionless form, as above. In this way, all formulas are expressed in terms of 
relative flows such as
\begin{equation}
 \kappa = \frac{I_1A_1}{I_2A_2}\,, \quad u_1 =\frac{A_1}{\widehat{Q}_1}\, , \quad u_2 =\frac{A_2}{\widehat{Q}_2} \, . 
\end{equation}
Parameters like
\begin{equation}
 \frac{I_2(\widehat{Q}_2-A_2)}{I_1A_1} \quad \mbox{and} \quad \frac{I_1(\widehat{Q}_1-A_1)}{I_2A_2}
\end{equation}
can be expressed through the previous set of parameters. A single intersection with 2
phases only is characterized by the 2 parameters $u_1$ and $u_2$, if queue minimization is performed, and one additional parameter $\kappa$, if travel time is minimized. Therefore, the optimal operation of $n$ intersections  depends on $2^n$ (or even $3^n$) parameters. In view of this, it is obvious that the optimal coordination of traffic lights in an urban road network constitutes a hard computational problem \cite{NPhard}.
\par
The consideration of non-uniform arrival flows further complicates matters. If the traffic flows are not constant, but characterized by vehicle platoons, the {\it phase} of traffic light control can be significant for intersection capacity \cite{EPJB8}. Therefore, the mutual coordination of neighboring traffic lights has a significant impact \cite{EPJB8}. This issue is, for example, addressed in Refs. \cite{Helbing2005,jstat}.

\section{Optimize-Multiple-Phases Approach}\label{MUL}

Under certain circumstances, it may be reasonable to {\it interrupt} the service of a vehicle queue to clear the way for a large flow of newly arriving vehicles in the other road section. Such an interruption may be interpreted as another slower-is-faster effect, occuring in situations where the interruption-induced delay of vehicles in one road section is overcompensated for by the avoidance of delay times in the other road section. Such effects involving several green phases can clearly not be studied within the optimization of a single phase. One would rather need an approach that optimizes two or more phases simultaneously. 
\par
In the optimize-two-phases approach, it appears logical to optimize the goal function
\begin{equation}
G_{12}(\tau_1 + \Delta T_1 + \tau_2 + \Delta T_2) = 
\frac{F_{\rm 12}(\tau_1 + \Delta T_1 + \tau_2 + \Delta T_2)} 
{\tau_1 + \Delta T_1 + \tau_2 + \Delta T_2} \, , 
\end{equation} 
which considers the waiting times in the successive green phase $\Delta T_2$ as well.
The average delay time $G_{\rm 12}(\tau_1 + \Delta T_1 + \tau_2 + \Delta T_2)$ is minimized by variation of {\it both} green time periods, $\Delta T_1$ and $\Delta T_2$. The optimal green times are characterized by vanishing partial derivatives $\partial G_{\rm 12}/\partial \Delta T_j$. Therefore, we must find those values $\Delta T_1$ and $\Delta T_2$ which fulfil
\begin{equation}
 \frac{\partial G_{\rm 12}}{\partial \Delta T_j}
= \frac{\displaystyle\frac{\partial F_{\rm 12}}{\partial \Delta T_j} (\tau_1+\Delta T_1 + \tau_2 + \Delta T_2) - F_{\rm 12}}{(\tau_1 + \Delta T_1 + \tau_2 + \Delta T_2)^2} = 0 \, .
\label{solv}
\end{equation}
This implies the balancing principle 
\begin{equation}
 \frac{\partial F_{\rm 12}(\tau_1+\Delta T_1 + \tau_2 + \Delta T_2) }{\partial \Delta T_1} 
 = \frac{\partial F_{\rm 12}(\tau_1+\Delta T_1 + \tau_2 + \Delta T_2) }{\partial \Delta T_2}  
\label{PRESS}
\end{equation}
which is known from other optimization problems as well, e.g. in economics \cite{Feichtinger}. Condition (\ref{PRESS}) allows one to express the green time $\Delta T_2$ as a function of the green time $\Delta T_1$. Both values can then be fixed by finding minima of $G_{\rm w}(\tau_1+\Delta T_1 + \tau_2 + \Delta T_2(\Delta T_1))$.
When this optimization procedure is applied after completion of each phase, it is expected to be adaptive to changing traffic conditions. However, a weakness of the above approach is its neglection of the flows in the optimization procedure. Therefore, the resulting intersection throughput may be poor, and flows would not necessarily be served, when the intersection capacity would allow for this.
Therefore, we will now modify the multiple-phase optimization in a suitable way, focussing on the two-phase case.

\subsection{Combined Flow-and-Delay Time Optimization}\label{Mixed}

The new element of the following approach is the introduction of flow constraints into the formulation of the delay time minimization. For this, let us start with the formula for the average delay time ${\cal T}_j^{\rm av}$ in road section $j$ derived in Ref. \cite{EPJB8}. It reads
\begin{equation}
 {\cal T}_j^{\rm av} = \frac{(1-f_j)^2}{(1-u_j)} \frac{T_{\rm cyc}}{2}
\end{equation}
with 
\begin{equation}
T_{\rm cyc} = \tau_1 + \Delta T_1 + \tau_2 + \Delta T_2 = (\tau_1+\tau_2)(1+\sigma_1+\sigma_2)
\end{equation} 
and 
\begin{equation}
1-f_j = \frac{T_{\rm cyc} - \Delta T_j}{T_{\rm cyc}} = \frac{(1+\sigma_1 + \sigma_2) - \sigma_j}{1+\sigma_1+\sigma_2} \, .
\end{equation}
As the number of vehicles arriving on road section $j$ during the time period $T_{\rm cyc}$ is given by
$I_j A_jT_{\rm cyc} =I_j u_j\widehat{Q}_jT_{\rm cyc}$, the average delay time of vehicles over the two green phases $\Delta T_1, \Delta T_2$ and amber time periods $\tau_1, \tau_2$ covered by the cycle time $T_{\rm cyc}(\Delta T_1,\Delta T_2)$ is given by
\begin{eqnarray}
G 
&=& \frac{\displaystyle\sum_{j=1}^2{\cal T}_j^{\rm av}I_j u_j\widehat{Q}_jT_{\rm cyc}}{T_{\rm cyc}} \nonumber \\
&=& \sum_{j=1}^2 \frac{[(1+\sigma_1+\sigma_2)-\sigma_j]^2}{2(1-u_j)(1+\sigma_1+\sigma_2)} I_j u_j\widehat{Q}_j (\tau_1+\tau_2) \, .  \qquad \label{THIS}
\end{eqnarray}
Let us now set $\theta_j = \theta_j(\sigma_1,\sigma_2) = 0$, if $\sigma_j \le \hat{\sigma}_j$ (corresponding to $\sigma_j/(1+\sigma_1+\sigma_2) \le u_j$), and $\theta_j = 1$ otherwise. The dimensionless clearing time
\begin{equation}
\hat{\sigma}_j  = \frac{u_j}{(1-u_j)} (1+\sigma_1 + \sigma_2 - \sigma_j) \label{RES11}
\end{equation}
was defined in Eq. (\ref{RES1}). With this, 
we will minimize the scaled average delay time  (\ref{THIS}) in the spirit of the optimize-two-cycles approach, but under the constraint that the average outflow 
\begin{eqnarray}
\overline{O} \!&=&  \! \sum_{j=1}^2 \frac{I_j \widehat{Q}_j \big\{ \Delta T_j (1-\theta_j) + [T_j + u_j(\Delta T_j - T_j)]\theta_j \big\}} {T_{\rm cyc}} \nonumber \\
&=& \!\sum_{j=1}^2 \!\frac{I_j \widehat{Q}_j \big\{ \sigma_j (1-\theta_j) 
+ [(1-u_j)\hat{\sigma}_j  + u_j \sigma_j ] \theta_j\big\}} 
{1 + \sigma_1+\sigma_2} \quad
\end{eqnarray}
reaches the maximum throughput 
\begin{equation}
\widehat{O}(u_1,u_2) = \min \Big( G_{\rm t}(u_1,u_2), O_{\rm max}(u_1,u_2)\Big) \, . 
\end{equation}
The maximum throughput corresponds to the overall flow $G_{\rm t}(u_1,u_2) = I_1A_1+I_2A_2 = u_1I_1\widehat{Q}_1 + u_2I_2\widehat{Q}_2$, as long as the capacity constraint (\ref{DIFF}) is fulfilled. 
Otherwise,  if the sum of arrival flows exceeds the intersection capacity, the maximum throughput is given by\footnote{If the cycle time $T_{\rm cyc}$ is limited to a certain maximum value $T_{\rm cyc}^{\rm max}$, one must replace the constraint $x_1+x_2 \le 1$ by $x_1+x_2 \le 1 - (\tau_1+\tau_2)/T_{\rm cyc}^{\rm max}$ and $1-u_2$ by $1 - u_2 - (\tau_1+\tau_2)/T_{\rm cyc}^{\rm max}$.} 
\begin{eqnarray}
O_{\rm max}(u_1,u_2) &=& \max_{x_j \le u_j \atop x_1+x_2 = 1}  \!\! \big( x_1 I_1\widehat{Q}_1 + x_2 I_2\widehat{Q}_2 \big)  \\
&=& \max_{1-u_2 \le x_1 \le u_1}  \!\! I_2\widehat{Q}_2 \big[ K x_1+ (1-x_1) \big] \nonumber \\
&=& \left\{
\begin{array}{ll}
I_2\widehat{Q}_2 \big[ (K-1)u_1 + 1 \big] &\mbox{if } K \ge 1 \\
I_2\widehat{Q}_2 \big[1 - (1-K)(1-u_2)\big]  & \mbox{if } K < 1.
\end{array}
\right.\quad \nonumber
\end{eqnarray}
Demanding the flow constraint
\begin{equation}
\overline{O}\Big(\sigma_1(u_1,u_2),\sigma_2(u_1,u_2)\Big) 
= \widehat{O}(u_1,u_2)    \label{fina1}
\end{equation}
and considering Eq. (\ref{RES11}), we can derive
\begin{equation}
\widehat{O} = \sum_j I_j\widehat{Q}_j \left[ \frac{\sigma_j(1-\theta_j)}{1+\sigma_1+\sigma_2}+ u_j\theta_j \right] \, . 
\end{equation}
This implies a linear relationship between $\sigma_1$ and $\sigma_2$. If the denominator is non-zero, we have:
\begin{eqnarray}
\sigma_1(\sigma_2)&=& \frac{\theta_1u_1I_1 \widehat{Q}_1  + \theta_2 u_2 I_2 \widehat{Q}_2  - \widehat{O}}{\widehat{O} - (1-\theta_1)I_1 \widehat{Q}_1  -  \theta_1 u_1 I_1\widehat{Q}_1 - \theta_2 u_2 I_2 \widehat{Q}_2  } \nonumber \\ 
& +& \frac{(1-\theta_2)I_2 \widehat{Q}_2  + \theta_1 u_1 I_1 \widehat{Q}_1+ \theta_2 u_2 I_2 \widehat{Q}_2 - \widehat{O}}{\widehat{O} - (1-\theta_1) I_1 \widehat{Q}_1  - \theta_1 u_1 I_1 \widehat{Q}_1  - \theta_2 u_2 I_2 \widehat{Q}_2 } \, \sigma_2 . \nonumber \\
 & & \label{sig12}
\end{eqnarray}
By demanding the flow constraint, we can guarantee that all arriving vehicles are served as long as the intersection capacity is sufficient, while we will otherwise use the maximum possible intersection capacity.
As a consequence, operation regime (0) of the one-phase optimization, which neglected the service of at least one road section, cannot occur within this framework. Instead, it is replaced by an operation regime, in which the vehicle queue in one road section is fully cleared, while the vehicle queue in the other road section is served in part.\footnote{In this case, we do not expect a periodic signal control anymore, as the growing vehicle queue in one of the road sections, see Ref. \cite{EPJB8}, has to be considered in the signal optimization procedure. Our formulas for one-phase optimization can handle this case due to the dependence on $\Delta N_j(0)$. In the two-phase optimization procedure, we would have to add $\sum_j I_j \, \Delta N_j(0)$ to formula (\ref{THIS}), where $\Delta N_j(0) = A_jT_{\rm cyc}^{k} - \widehat{Q}_j\,\Delta T_j^{k}$ denotes the number of vehicles that was not served during the $k$th cycle $T_{\rm cyc}^k = \tau_1+\Delta T_1^k+\tau_2+\Delta T_2^k$. This gives an additional term $\sum_j u_jI_j\widehat{Q}_j (\tau_1+\tau_2) \sum_k (1+\sigma_1^k +\sigma_2^k - \sigma_j^k/u_j)$ in Eq. (\ref{THIS}).} Of course, this will happen only, if the intersection capacity is insufficient to serve both flows completely (i.e. in the case $1-u_1-u_2 < 0$). If $K>1$ (i.e. the main flow is on road section 1), we have
\begin{equation}
 \frac{\sigma_1}{1+\sigma_1+\sigma_2} = u_1 \quad \mbox{and} \quad
 \frac{\sigma_2}{1+\sigma_1+\sigma_2} =  (1-u_1) \, .   
\end{equation}
If $K<1$, the indices 1 and 2 must be interchanged. 
\par
\begin{figure*}[htbp]
\begin{center}
\hbox{\includegraphics[width=9.2cm]{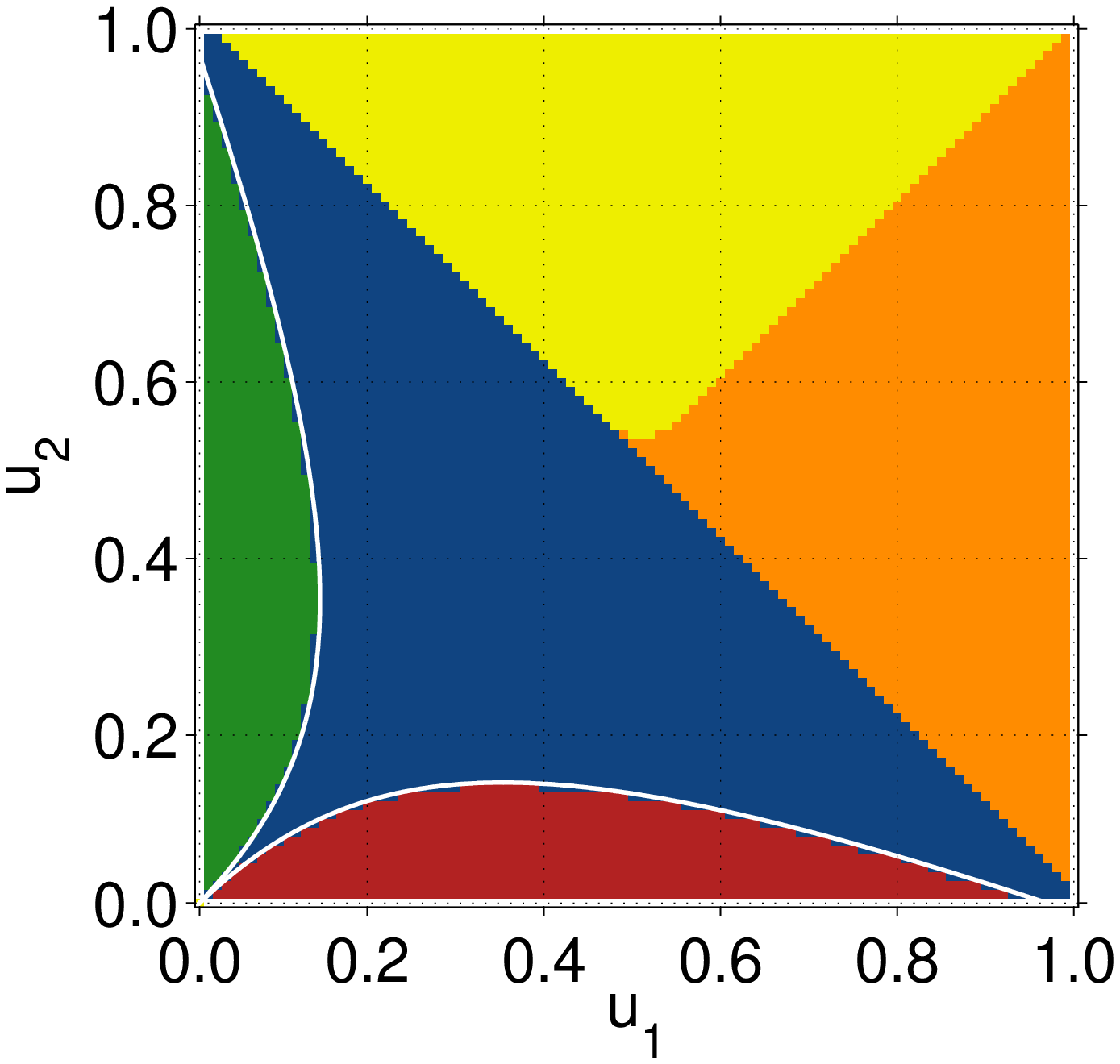}
\includegraphics[width=9.2cm]{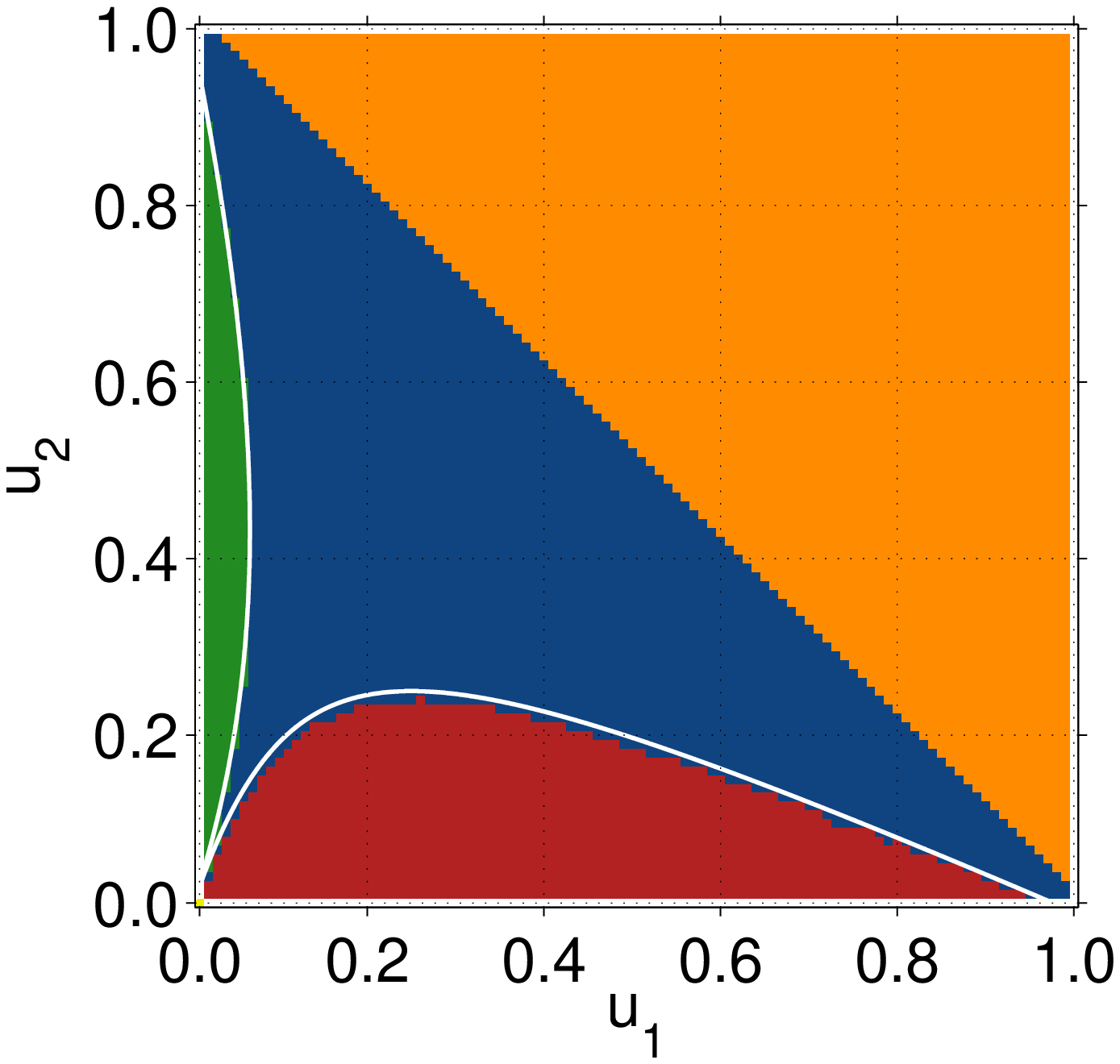}}
\end{center}
\caption[]{Operation regimes of periodic signal control as a function of the utilizations $u_j$ of both road sections according to the  two-phase optimization approach, assuming $K=1$, corresponding to equal roads (left), and $K=3$, corresponding to a three-lane road 1 and a one-lane road 2 (right). 
For most combinations of utilizations (if $u_1$ is not too different from $u_2$), the green phases are terminated as soon as the corresponding road sections are cleared (see the blue area below the falling diagonal line). However, extended green times for road section 1 result (see the red area along the $u_1$ axis), if the utilization of road section 2 is small. In contrast, if the utlization of road section 1 is small, extended green times should be given to road section 2 (see the green area along the $u_2$ axis). The white separating lines between these areas correspond to Eqs. (\ref{SepLines}), (\ref{SepLin}) fit the numerical results well.  Above the line $u_2 = 1-u_1$, the intersection capacity is insufficient to serve the vehicle flows in {\it both} road sections. In this area, the two-phase optimization gives solutions where road section 1 is fully cleared, but road section 2 is served in part (orange area towards the right), or vice versa (yellow area towards the top in the left figure).}
\label{displ5}
\end{figure*}

Operation regime (1) is still defined as in Sec. \ref{peropreg} and characterized by
\begin{equation}
 \sigma_j = \frac{u_j}{1-u_1-u_2} \, , \qquad \frac{\sigma_j}{1+\sigma_1+\sigma_2} = u_j \, . 
\label{still}
\end{equation}
In contrast to the one-phase optimization approach, this ``normal case'' of signal operation occurs in a large parameter area of the two-phase optimization approach (see blue area in Fig. \ref{displ5}). It implies that both green times are long enough to dissolve the vehicle queues, but not longer.
\par
\begin{figure*}[htbp]
\begin{center}
\hbox{\includegraphics[width=9.2cm]{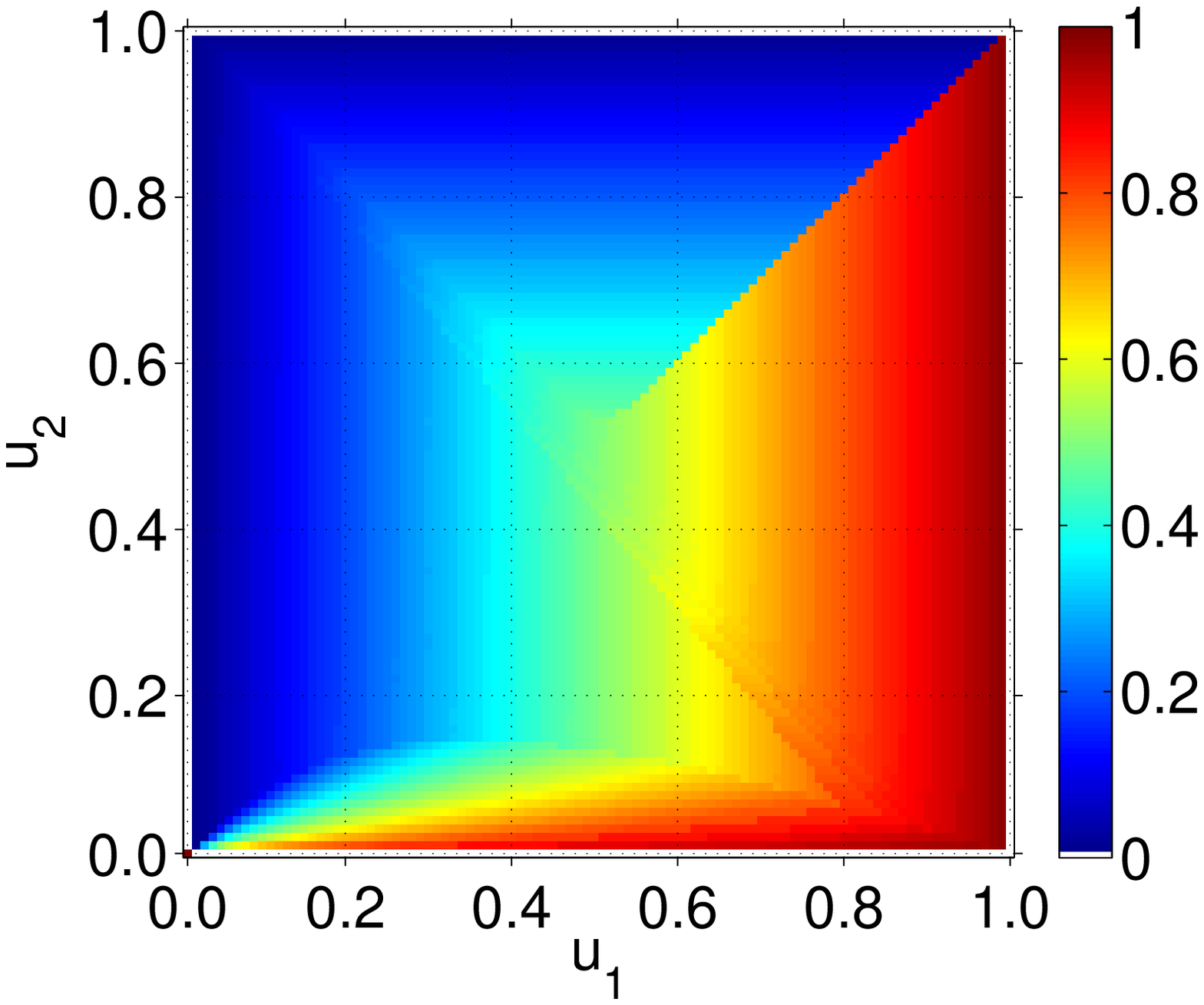}
\includegraphics[width=9.2cm]{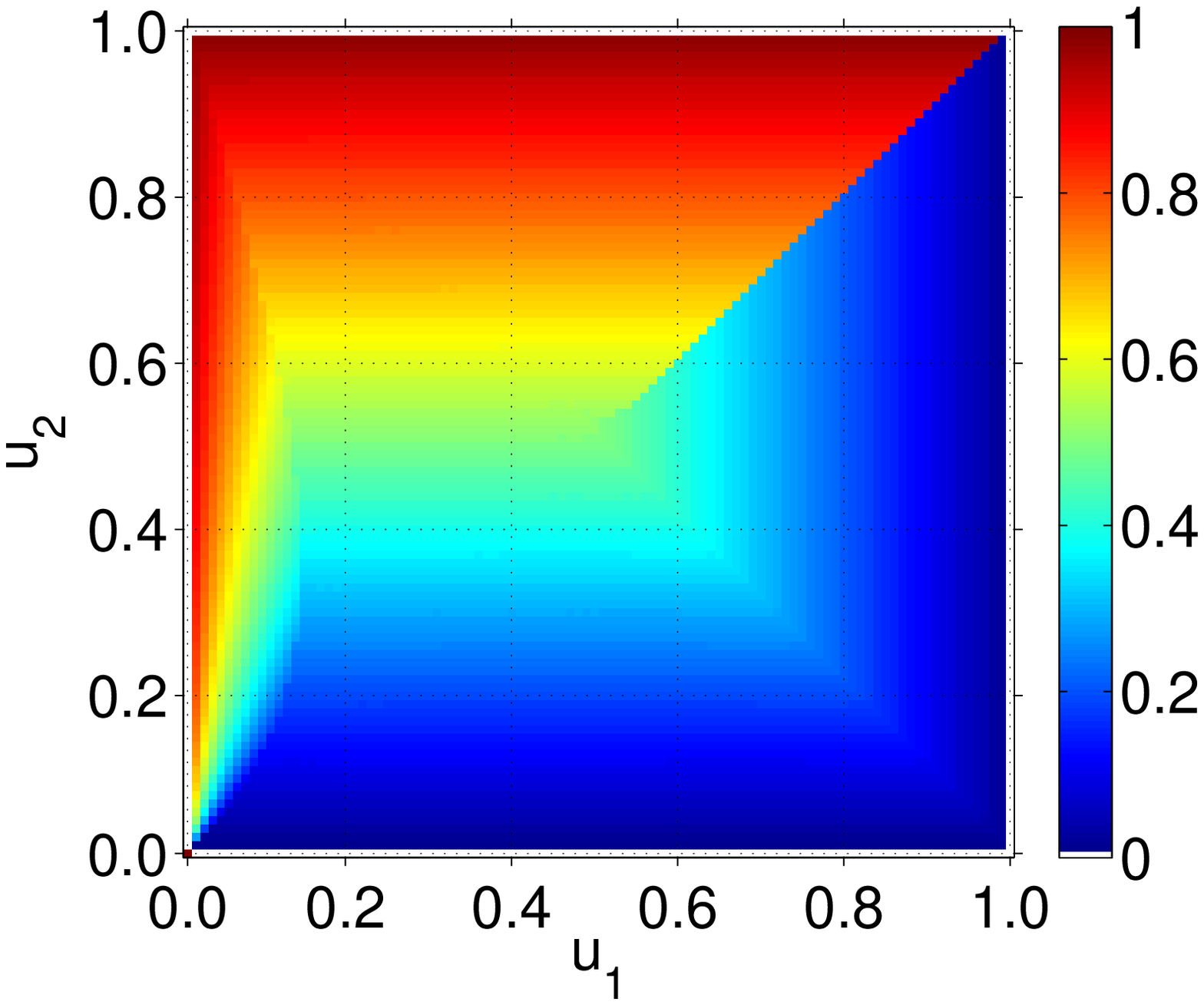}}
\end{center}
\caption[]{Optimal green time fractions $\Delta T_j /T_{\rm cyc} = \sigma_j / (1+\sigma_1+\sigma_2)$ for road section $j=1$ (left) and road section $j=2$ (right) as a function of the utilizations $u_j$ of both roads $j$, assuming periodic signal operation according to the two-phase optimization approach with $K=1$. For combinations $(u_1,u_2)$ with several solutions (with extended green time and without), we display the solution which minimizes the goal function (\ref{THIS}). The results are qualitatively similar to the ones belonging to the one-phase optimization approach displayed in Fig. \ref{pha1}, but we find periodic solutions above the capacity line $u_2 = 1-u_1$, where one road section (the one with the greater utilization) is fully cleared, while the other one is served in part.}\label{displ3}
\end{figure*}
\begin{figure*}[htbp]
\begin{center}
\hbox{\includegraphics[width=9.2cm]{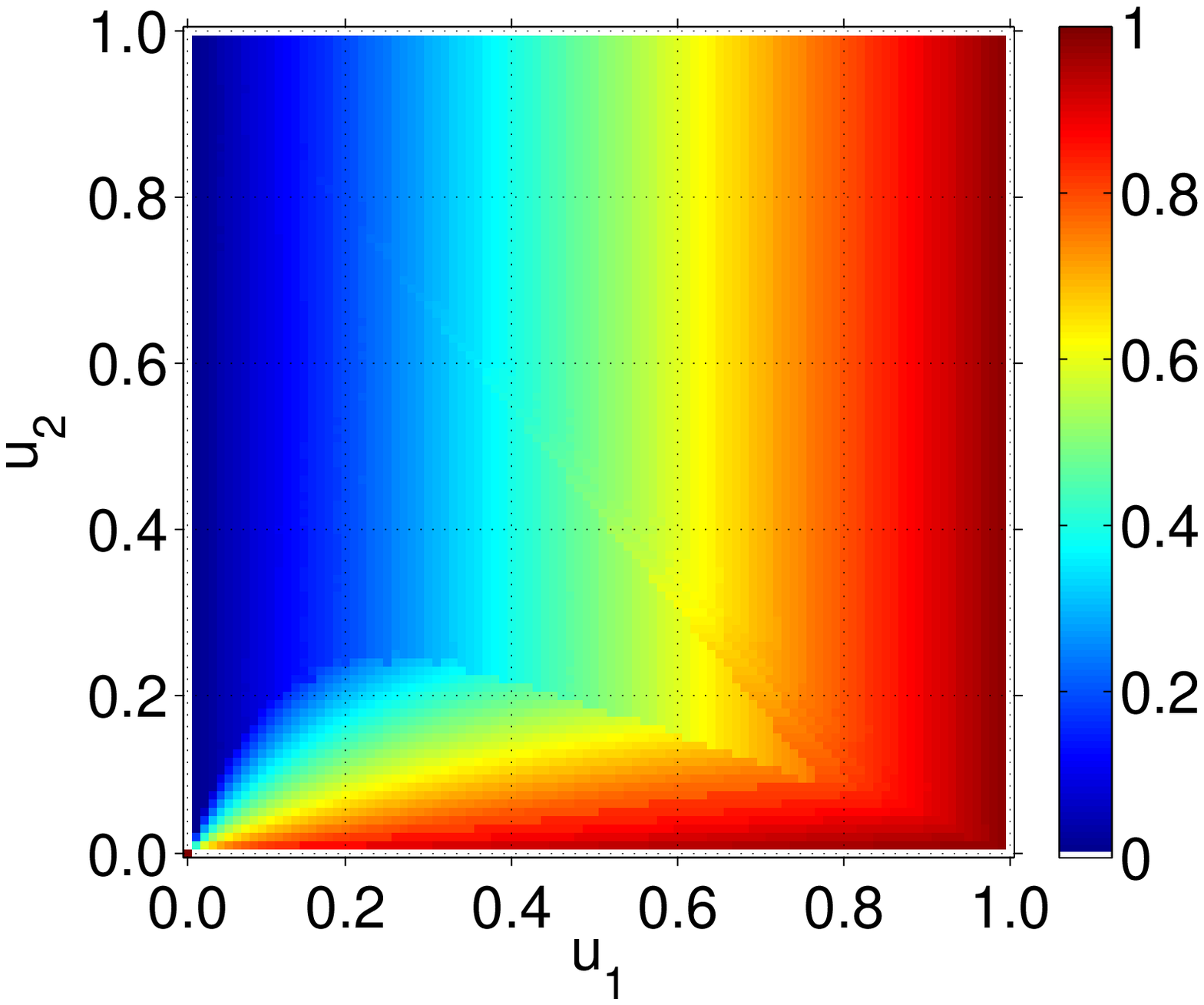}
\includegraphics[width=9.2cm]{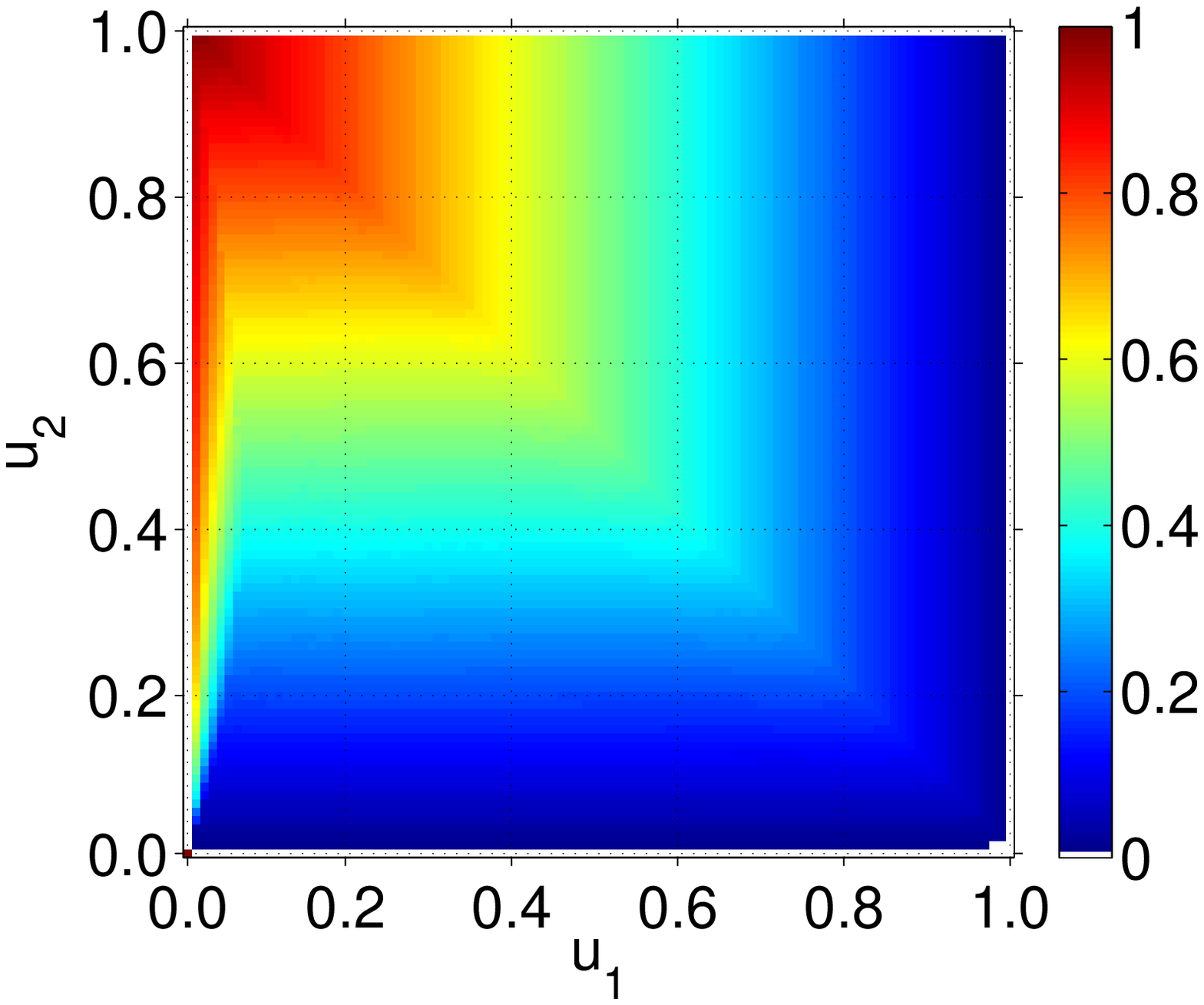}}
\end{center}
\caption[]{Same as Fig. \ref{displ3}, but for $K=3$, corresponding to a three-lane road section 1 (arterial road) and a one-lane road section~2 (crossing side road).}\label{displ4}
\end{figure*}

The case, where both green phases are extended, is again no optimal solution. We will, therefore, finally focus on case (2), where the vehicle queue in road section 1 is just cleared ($\theta_1=0$), while road section 2 gets an excess green time ($\theta_2 = 1$). With $\widehat{O} = G_{\rm t}= u_1I_1\widehat{Q}_1 + u_2I_2\widehat{Q}_2$, Eq. (\ref{sig12}) yields the simple constraint
\begin{equation}
 \sigma_1(\sigma_2) = \frac{u_1}{1-u_1}(1+\sigma_2) \, , \label{insi}
\end{equation}
which corresponds to Eq. (\ref{RES1}). It implies
\begin{equation}
\frac{d\sigma_1}{d\sigma_2} = \frac{u_1}{1-u_1}\, , \qquad 1+\sigma_1 = \frac{1+u_1\sigma_2}{1-u_1}\, ,
\end{equation}
and
\begin{equation} 
 1+\sigma_1+\sigma_2 = \frac{1+\sigma_2}{1-u_1} \, , \qquad \frac{\sigma_1}{1+\sigma_1+\sigma_2} = u_1 \, .  
\end{equation}
We will now determine the minimum of the goal function $G$ by setting the derivative $\partial G/\partial \sigma_1$ to zero,  considering
\begin{equation}
\frac{d\hat{\sigma}_2(\sigma_1)}{d\sigma_1} = \frac{u_2}{1-u_2} \, . 
\end{equation}
Multiplying the result with $2(1-u_1)^3(1-u_2)(1+\sigma_1+\sigma_2)^2/(I_2\widehat{Q}_2)$, we find the following relationship:
\begin{eqnarray}
& & 2 u_1 u_2 (1+\sigma_2) (1+u_1\sigma_2) \nonumber \\
&+& 2Ku_1(1-u_1)(1-u_2)(1+\sigma_2)^2 \nonumber \\
&=& u_2(1+u_1\sigma_2)^2 + Ku_1(1-u_1)(1-u_2)(1+\sigma_2)^2 \, , \qquad 
\end{eqnarray}
which finally leads to
\begin{eqnarray}
(1+\sigma_2)^2 &=& \frac{u_2(1-u_1)^2}{u_1{}^2u_2 + Ku_1(1-u_1)(1-u_2)} \nonumber \\
&=& \frac{(1-u_1)^2}{u_1{}^2 + \kappa (1-u_1)(1-u_2)} \, . 
\label{spul}
\end{eqnarray}
According to Eq. (\ref{atisf}), for an extended green time on road section 2, the condition
\begin{equation}
\frac{\sigma_2}{1+\sigma_1+\sigma_2} > u_2 
\end{equation}
must again be fulfilled.
If the solution $\sigma_2(u_1,u_2)$ of Eq. (\ref{spul}) satisfies this requirement, it can be inserted into Eq. (\ref{insi}) to determine the scaled green time period $\sigma_1(u_1,u_2)$ as a function of the capacity utilizations $u_1$ and $u_2$  within the framework of the optimize-two-phases approach. The corresponding results are displayed in Figs. \ref{displ5} to \ref{displ4}. A generalization to signal controls with more than two phases is straightforward.
\par
Finally, let us calculate the separating line between case (1) and case (2). Inserting Eq. (\ref{insi}) into (\ref{THIS}), we can express the goal function $G$ as a function $H$ of a single variable $\sigma_2$:
\begin{equation}
H(\sigma_2) = G(\hat{\sigma}_1(\sigma_2),\sigma_2) \, . \label{Haa}
\end{equation}
As Eq. (\ref{insi}) holds for both cases, an exact clearing of road section 2 or an excess green time for it, 
the functional dependence of goal function (\ref{Haa}) on $\sigma_2$ must be the same for both cases. Now, on the one hand, we may apply Eq. (\ref{still}) for the case without excess green time, which yields
\begin{equation}
1+\sigma_2 = \frac{1-u_1}{1-u_1-u_2} \quad \mbox{and} \quad 
(1+\sigma_2)^2 = \frac{(1-u_1)^2}{(1-u_1-u_2)^2} \, .  
\end{equation}
On the other hand, in the case of excess green time, we may use Eq. (\ref{spul}). The goal function must be the same along the separating line between both cases, which requires
\begin{equation}
\frac{(1-u_1)^2}{(1-u_1-u_2)^2} = \frac{(1-u_1)^2}{u_1{}^2 + \kappa (1-u_1)(1-u_2)} \, . 
\end{equation}
This implies
\begin{equation}
 \kappa(1-u_1)(1-u_2) = (1-u_1-u_2)^2 - u_1{}^2
\end{equation}
or
\begin{equation}
\kappa(1-u_1)(1-u_2) = (1-2u_1-u_2)(1-u_2) \, . 
\end{equation}
The finally resulting equation for the separating line between the regimes with and without excess green time is given by
\begin{equation}
\frac{1}{\kappa} = \frac{u_2}{Ku_1} = \frac{1-u_1}{1-2u_1-u_2} \, . \label{SepLines}
\end{equation}
As Fig. \ref{displ5} shows, this analytical result fits the result of our numerical optimization very well.
The separating line between case (1) and case (3) is derived analogously. It may also be obtained by interchanging the subscripts 1 and 2 and substituting $\kappa$ by $1/\kappa$, yielding
\begin{equation}
\kappa = K \frac{u_1}{u_2}  = \frac{1-u_2}{1-2u_2 - u_1} \, . \label{SepLin}
\end{equation} 

\section{Summary, Discussion, and Outlook}\label{sUm}

\begin{figure}[htpb]
\centering
\includegraphics[width=9.2cm]{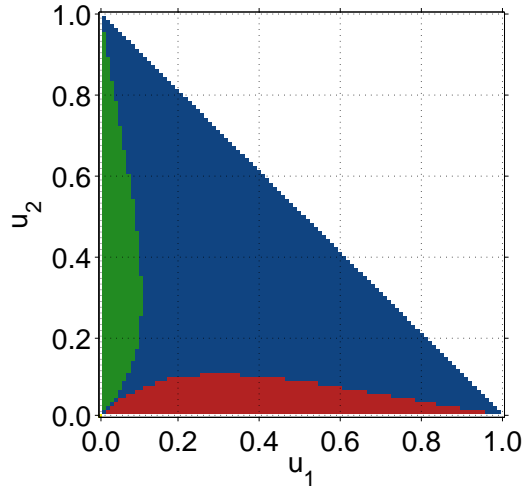}
\caption[]{Operation regimes of periodic signal control as a function of the utilizations $u_j$ of both roads,
if one specifies the clearing times and excess green times according to Eqs. (\ref{RES1}) and (\ref{RES2}) of the one-phase optimzation, but selects the solution that minimizes the overall delay time (\ref{THIS}) over two successive phases. For most combinations of utilizations (if $u_1$ is not too different from $u_2$), the green phases are terminated as soon as the corresponding road sections are cleared (see the blue area below the falling diagonal line). However, extended green times for road section 1 result (see the red area along the $u_1$ axis), if the utilization of road section 2 is small. In contrast, if the utilization of road section 1 is small, extended green times should be given to road section 2 (see green area along the $u_2$ axis) \cite{StRef}. Above the line $u_2 = 1-u_1$, the intersection capacity is insufficient to serve the vehicle flows in {\it both} road sections.}\label{pha1}
\end{figure}

We have studied the control of traffic flows at a single intersection. Such studies have been performed before, but we have focussed here on some particular features:
\begin{itemize}
\item For the sake of a better understanding, we were interested in deriving analytical formulas, even though this required some simplifications.
\item A one-phase minimization of the overall travel times in all road sections tended to give excess green times to the main flow, i.e. to the road section with the larger number of lanes or, if the number of lanes is the same ($K=1$), to the road section with the larger utilization (see Fig. \ref{newfig}). The excess green time can lead to situations where one of the vehicle flows is not served, although there would be enough service capacity for all flows.
\item A minimization of vehicle queues rather than travel times simplifies the relationships through the special settings $\widehat{Q}_j = \widehat{Q}$ and $I_j = 1$, resulting in $K=1$. Moreover, these settings guarantee that the case of no service only occurs, if the intersection capacity is exceeded.
\item An optimize-multiple-phases approach considering flow constraints gives the best results among the optimization methods considered. It makes sure that both roads are served even when the intersection capacity is exceeded. 
\item For all considered optimization approaches, we have derived different operation regimes of traffic signals control: One of them is characterized by ending a green time period upon service of the last vehicle in the queue, which implies that all vehicles are stopped once by a traffic signal. However, we have also found conditions under which it is advised to delay switching for one of the road sections (``slower-is-faster effect''), which allows some vehicles to pass the signal without stopping.
\item Compared to the one-phase optimization, a two-phase optimization tends to to give much less excess green times, in particular if the utilizations of the road sections are comparable. We hypothesize that this is an effect of the short-sightedness of the one-phase optimization: It does not take into account future delay times caused by current excess green times. This hypothesis is confirmed by Fig. \ref{pha1}
(which is to be contrasted with the left illustration in Fig. \ref{newfig}). It specifies the green time durations according to Eqs. (\ref{RES2}) and (\ref{RES1}) of the one-phase optimzation, but selects the solution that minimizes the average delay time (\ref{THIS}) over two phases.
\item Although the multi-phase optimization approach provides extended green times in a considerably smaller area of the parameter space spanned by the utilizations $u_j$, the slower-is-faster effect still persists when signal settings are optimized over a full cycle time (as we effectively did with the periodic two-phase optimization approach). The slower-is-faster effect basically occurs when the utilization of a road section is so small that it requires some extra time to collect enough vehicles for an efficient service during the green phase, considering the efficiency losses by switching traffic lights during the amber phases.
\item In complementary appendices, we discuss traffic controls with more than two phases and an exponentially weighted goal function for short-term traffic optimization.
Furthermore, we propose how to take into account the effect of stopping newly arriving vehicles and how to assess its impact as compared to queues of waiting vehicles. As stopping vehicles causes additional delay times, it becomes often favorable to implement excess green times (i.e. to apply the slower-is-faster effect''). 
\end{itemize}
Our restriction to analytical calculations implied certain simplifications such as the assumption of two traffic phases,  the assumption of constant arrival flows, and no obstructions of the outflow. However, these restrictions can be easily overcome by straight-forward generalizations (see Appendices). The assumption of constant arrival flows, for example, is not needed. Assuming a short-term prediction based on upstream flow measurements \cite{anticip}, the expected delay times or queue lengths can be determined via the integral (\ref{instead}). The optimal solution must then be numerically determined, which poses no particular problems. Although the behavior may become somewhat more complicated and the boundaries of the operation regimes may be shifted, we expect that the above mentioned signal operation modes and the control parameters $u_1=A_1/\widehat{Q}_1$, $u_2=A_2/\widehat{Q}_2$, and $\kappa = I_1A_1/(I_2A_2)$ still remain relevant. 
\par
Finally, the above described traffic light control principle can be generalized to 
cases where the outflow from a road section during a green phase is blocked due to spillover effects. As this implies growing delay times in this road section {\it and} all the others, travel time minimization in this case will interrupt the service in favor of a road section that can be successfully left by vehicles when a green light is given to them. In summary, our approach successfully delivers analytical insights into various operation regimes of traffic signal control, including the occuring slower-is-faster effects. Moreover, as the two-phase optimization approach takes care of side roads and minor flows, it has similar effects as the stabilization rule that was introduced in Ref. \cite{jstat} to compensate for unstable service strategies. 

\begin{acknowledgement}
{\it Author contributions:} DH set up the model, performed the analytical calculations, and wrote the manuscript. AM produced the numerical results and figures, and derived the separating lines given by 
Eqs. (\ref{SepLines}) and (\ref{SepLin}).\\
{\it Acknowledgements:} The authors would like to thank for partial support by the ETH project CH1-01 08-2 
the VW Foundation Project I/82 697, 
the Daimler-Benz Foundation Project 25-01.1/07, and 
the NAP project KCKHA005.

\end{acknowledgement}

\appendix

\section{Considering the Price of Stopping Vehicles}\label{plato}

The previous considerations have only taken into account delays by vehicles in a vehicle queue. 
However, it would also make sense to consider the 
price of stopping vehicles. In particular, it must be possible that a large flow of moving vehicles in one road section is prioritized to a short queue of standing vehicles in the other road section. But 
how can we assess the relative disadvantage of stopping newly arriving vehicles as compared to stopping the service of a vehicle queue at the intersection? If the arrival flow is not large enough, it would certainly be better to continue serving the standing vehicle queue in the other road until it is fully dissolved.
\par
We pursue the following approach: While the flow model used before implicitely assumes instantaneous
vehicle accelerations and decelerations, we will now consider that, in reality, a finite vehicle acceleration $a$ causes additional delays of $V_j^0/(2a)$, where $V_j^0$ denotes the free speed or speed limit. Furthermore, 
the reaction time $T_{\rm r}$ must be taken into account as well. This leads to an additional delay of
\begin{equation}
 T'_j = T_{\rm r} + \frac{V_j^0}{2a} 
\end{equation}
for each vehicle that leaves a queue. $T_{\rm r}$ is of the order of the safe 
time gap $T$. Note that delays $V_j^0/(2b)$ due to a finite deceleration $b$ do
not additionally contribute to the delay times, as it does not matter whether 
delayed vehicles spend their time decelerating or stopped.\footnote{The finite deceleration
only matters slightly, when the exact moment must be determined when a road
section becomes fully congested.}
\par
Furthermore, we must determine the rate at which such additional delays are produced. This is given
by the rate at which freely moving vehicles join the end of a traffic jam, i.e. by
\begin{equation}
\rho_{\rm jam} |C_j| = \frac{ \rho_{\rm jam}}{\rho_{\rm jam}/A_j - 1/V_j^0} \ge A_j \, , 
\end{equation}
where $\rho_{\rm jam}$ denotes the density of vehicles per lane in a standing queue.
The propagation speed 
\begin{equation}
C_j = \frac{A_j-0}{A_j/V_j^0 - \rho_{\rm jam}} 
\end{equation}
of the upstream front of the queue
corresponds to the propagation speed of shock fronts, see Refs. \cite{wHitham,Helbing2003a,Helbing2007}. 
Depending on the values of $C_j$ (or $A_j$) and $T'_j$, newly arriving vehicles can have an impact $T'_jC_j\rho_{\rm jam}$ equivalent to about $\Delta N_j =10$ queued vehicles.
\par
Summarizing the above considerations, we suggest to replace the goal function $G_1(t)$ by the generalized formula 
\begin{equation}
 \widehat{G}_1(t) 
 = \frac{1}{t} \sum_j I_j \!\! \int\limits_0^t \! dt' \Big[ \Delta N_j(t') 
+  T'_j |C_j| \rho_{\rm jam} \Theta(\Delta N_j>0) \Big] \, ,
\label{instead1}
\end{equation}
where $\Theta(\Delta N_j>0)=1$, if $\Delta N_j >0$, and $\Theta(\Delta N_j>0)=0$ otherwise. 
In case (a) with $\Delta T_i \le T_i$, we find
\begin{eqnarray}
 \widehat{F}_1^{\rm a}(\tau_1+ \Delta T_i+\tau_2) &=& F_1^{\rm a}(\tau_1+\Delta T_i+\tau_2) \nonumber \\
 &+& I_1T'_1|C_1|\rho_{\rm jam} (\tau_1+\Delta T_1+\tau_2) \nonumber \\
 &+& I_2T'_2|C_2|\rho_{\rm jam} (\tau_1+\Delta T_1+\tau_2) \, .  \qquad
\end{eqnarray}
This implies
\begin{eqnarray}
 \widehat{G}_1^{\rm a}(\tau_1+\Delta T_i+\tau_2) &=& G_1^{\rm a}(\tau_1+\Delta T_i+\tau_2) \nonumber \\
 &+& I_1T'_1|C_1|\rho_{\rm jam} + I_2T'_2|C_2|\rho_{\rm jam} \qquad
 \label{Ga1}
\end{eqnarray}
with $G_1^{\rm a}(\tau_1+\Delta T_i+\tau_2)$ according to Eq. (\ref{GAIN1}).
Therefore, the partial derivative of $\widehat{G}_1^{\rm a}(\tau_1+\Delta T_i+\tau_2)$ with respect to $\Delta T_1$ remains unchanged, and we find the same optimal green time period $\Delta T_1 = 0$ or $\Delta T_1 \ge T_1$.
However, in case (b) with $\Delta T_1 \ge T_1$, we obtain
\begin{eqnarray}
 \widehat{F}_1^{\rm b}(\tau_1+ \Delta T_i+\tau_2) &=& F_1^{\rm b}(\tau_1+\Delta T_i+\tau_2) \nonumber \\
 &+& I_1T'_1|C_1|\rho_{\rm jam} (\tau_1+ T_1+\tau_2) \nonumber \\
 &+& I_2T'_2|C_2|\rho_{\rm jam} (\tau_1+\Delta T_1+\tau_2) \, ,  \qquad 
\end{eqnarray}
which implies
\begin{eqnarray}
 \widehat{G}_1^{\rm b}(\tau_1+\Delta T_i+\tau_2) &=& G_1^{\rm b}(\tau_1+\Delta T_i+\tau_2) \nonumber \\
 &+& I_1T'_1|C_1|\rho_{\rm jam} + I_2T'_2C_2\rho_{\rm jam} \nonumber \\ 
 &-&  I_1T'_1|C_1|\rho_{\rm jam}\frac{\Delta T_1-T_1}{\tau_1+\Delta T_1+\tau_2} \qquad  
 \label{Ga2}
\end{eqnarray}
with $G_1^{\rm b}(\tau_1+\Delta T_i+\tau_2)$ according to Eq. (\ref{GAIN2}).
In cases where an excess green time is favorable, the corresponding formula for the green time duration becomes
\begin{equation}
 (\tau_1+\Delta T_1+\tau_2)^2 = \frac{2I_1}{I_2A_2} [ E_1 + T'_1|C_1|\rho_{\rm jam} (\tau_1 + T_1 + \tau_2)] \, ,
\end{equation}
i.e. the optimal green times tend to be longer.
In order to support excess green times, the condition $ (\tau_1 + \Delta T_1+\tau_2)^2 \ge
(\tau_1+ T_1+\tau_2)^2$ must again be fulfilled, which requires
\begin{eqnarray}
 & &  \frac{(\Delta N_1^{\rm max})^2}{\widehat{Q}_1 - A_1}  \left( \frac{I_1}{I_2A_2} - \frac{1}{\widehat{Q}_1 - A_1}\right) \nonumber \\
 & & +2 \Delta N_1^{\rm max} \left( \frac{I_1\tau_1}{I_2 A_2} - \frac{\tau_1+\tau_2}{\widehat{Q}_1-A_1}\right) 
 \nonumber \\
&\ge & (\tau_1+\tau_2)^2 - \frac{2I_1T'_1|C_1|\rho_{\rm max} }{I_2A_2}  \left( \tau_1 + \frac{\Delta N_1^{\rm max}}{\widehat{Q}_1-A_1} + \tau_2\right) \, .
 \nonumber \\
& & 
\label{ro}
\end{eqnarray}
Comparing this with formula (\ref{verysmall}), we can see that the threshold for the implementation of excess green times $\Delta T_j > T_j$ is reduced. Therefore, excess green times will be implemented more frequently,  as this reduces the number of stopped vehicles. 

\section{More than Two Traffic Phases}\label{Be}

The above formulas for the optimize-one-phase approach can be easily generalized to multiple traffic phases of more complicated intersections as in the case of Barcelona's center (see Fig. \ref{Illx}). For
\begin{equation}
\Delta T_i \le T_i = \frac{\Delta N_i^{\rm max}}{\widehat{Q}_i - A_i}  \label{forexample}
\end{equation}
with 
\begin{equation}
\Delta N_i^{\rm max} = \Delta N_i(0) + A_i \tau_i\, , 
\label{noch1}
\end{equation}
for example, we can derive from Eq. (\ref{Ga1}) 
\begin{eqnarray}
& &  \widehat{G}_i^{\rm a}(\tau_i + \Delta T_i + \tau_{i+1}) \nonumber \\
&=& I_i \bigg[ \Delta N_i(0) + \widehat{Q}_i \tau_{i+1} \nonumber \\
 & & - (\widehat{Q}_i-A_i) \frac{\tau_i + \Delta T_i+\tau_{i+1}}{2} \bigg] \nonumber \\
&+& \sum_{j(\ne i)} I_j \left[ \Delta N_j(0) + A_j \frac{\tau_i + \Delta T_i+ \tau_{i+1}}{2} \right]  \nonumber \\
 &+& \sum_j I_jT'_j|C_j|\rho_{\rm jam} \, . 
 \label{AAA1}
\end{eqnarray}
In contrast, for $\Delta T_i \ge T_i$ and with 
\begin{equation}
E_i = \Delta N_i^{\rm max} \tau_i + \frac{(\Delta N_i^{\rm max})^2}{2(\widehat{Q}_i-A_i)} \, , 
\label{noch2}
\end{equation}
from Eqs. (\ref{Ga2}) and (\ref{GAIN2}) we obtain
\begin{eqnarray}
 & & \widehat{G}_i^{\rm b}(\tau_i+\Delta T_i + \tau_{i+1}) \nonumber \\
 &=& \frac{I_iE_i}{\tau_i+\Delta T_i + \tau_{i+1}} \nonumber \\ 
&+& \sum_{j(\ne i)} I_j\left[ \Delta N_j(0) + A_j \frac{\tau_i + \Delta T_i + \tau_{i+1}}{2} \right] \nonumber \\
&+& \sum_j I_jT'_j|C_j|\rho_{\rm jam} -  I_iT'_i|C_i|\rho_{\rm jam}\frac{\Delta T_i-T_i}{\tau_i+\Delta T_i+\tau_{i+1}} \, .
\nonumber \\ & & 
\label{AAA2}
\end{eqnarray}
The minimum of this function is reached for
\begin{equation}
(\tau_i + \Delta T_i + \tau_{i+1})^2  
= \frac{I_iE_i + I_iT'_i|C_i|\rho_{\rm jam}(\tau_i + T_i + \tau_{i+1})}{\sum_{j(\ne i)} I_jA_j/2} \, . 
\label{naendlich}
\end{equation}
The occurence of excess green time requires $ (\tau_i + \Delta T_i+\tau_{i+1})^2 \ge
(\tau_i+ T_i+\tau_{i+1})^2$, i.e.
\begin{eqnarray}
 & &  \frac{(\Delta N_i^{\rm max})^2}{\widehat{Q}_i - A_i}  \left( \frac{I_i}{\sum_{j(\ne i)} I_jA_j} - \frac{1}{\widehat{Q}_i - A_i}\right) \nonumber \\
 & & +2 \Delta N_i^{\rm max} \left( \frac{I_i\tau_1}{\sum_{j(\ne i)} I_j A_j} - \frac{\tau_i+\tau_{i+1}}{\widehat{Q}_i-A_i}\right)  \nonumber \\
&\ge & (\tau_i+\tau_{i+1})^2 - \frac{2I_iT'_i|C_i|\rho_{\rm jam} }{\sum_{j(\ne i)}I_jA_j}  \left( \tau_i + \frac{\Delta N_i^{\rm max}}{\widehat{Q}_i-A_i} + \tau_{i+1}\right) \, .  \nonumber \\
& & 
\label{ron}
\end{eqnarray}
It can be seen that the existence of more traffic phases is unfavorable for providing excess green times. For their existence, a small number of phases is preferable.

\subsection*{Procedure of Traffic Signal Control}

Based on the above formulas, the next green phase $i$ is determined as follows: 
\begin{enumerate}
\item Set the time $t$ to zero, after the last green phase $i'$ has been completed.
\item Apply the required service time (amber time) of duration $\tau_{i'+1}$ and set $\tau_j = \tau_{i'+1}$ for all road sections $j$. Then, calculate $\Delta N_j^{\rm max}$ and $E_j$ for all $j$ with formulas (\ref{noch1}) and (\ref{noch2}).
\item During the service time, determine the green times $\Delta T_j$ and $T_j$ with and without green time extension, for each road section $j$ with formulas (\ref{naendlich}) and (\ref{forexample}). 
\item If $\Delta T_j > T_j$ and $\widehat{G}_j^{\rm b}(\tau_j+\Delta T_j +\tau_{j+1}) < \widehat{G}_j^{\rm a}(\tau_j+T_j +\tau_{j+1})$, see Eqs. (\ref{AAA2}) and (\ref{AAA1}), consider the implementation of the extended green time $\Delta T_j$ and set $\widehat{G}_j = \widehat{G}_j^{\rm b}(\tau_j+\Delta T_j +\tau_{j+1})$. Otherwise consider the implementation of the clearing time $T_j$ and set $\widehat{G}_j = \widehat{G}_j^{\rm a}(\tau_j+T_j +\tau_{j+1})$, but if $\widehat{G}_j^{\rm a}(\tau_j+\tau_{j+1}) < \widehat{G}_j$, set $\Delta T_j = 0$ and $\widehat{G}_j =\widehat{G}_j^{\rm a}(\tau_j+\tau_{j+1})$.
\item Among all road sections $j'$ different from the previously selected one $i'$, choose that one $i$ for service, for which the expected average travel time $\widehat{G}_i$ is smallest (i.e. $\widehat{G}_i = \min_{j(\ne i')} \widehat{G}_j$). Implement the selected green phase $\Delta T_i$.
\item Update the length of the vehicle queue in road section $i$ according to 
\begin{equation}
\Delta N_i(\tau_i+\Delta T_i) = 0
\end{equation} 
and the queue lengths in all other road sections $j\ne i$ according to
\begin{equation}
\Delta N_j(\tau_i+\Delta T_i) = \Delta N_j(0) + A_j(\tau_i+\Delta T_i) \, . \label{upda}
\end{equation} 
If road section was not served ($\Delta T_i = 0$), update the vehicle queues in {\it all} road sections $j$ (including $i$) according to Eq. (\ref{upda}). 
\item At the end of the corresponding green time duration $\Delta T_i$, set $i' = i$ and continue with step 1. 
\end{enumerate}
The optimize-multiple-phases approach can be generalized in a similar way. Then, among all solutions satisfying preset flow constraints, that multi-phase solution is chosen, which  minimizes the goal function and does not start with a service of the previously served road section. In order to flexibly adjust to varying traffic conditions, one may repeat the optimization after completion of one phase rather than after completion of all the phases considered in the multi-phase optimization.

\section{Limited Forecast Time Horizon}\label{Ce}

While traffic light optimization is an NP-hard problem \cite{NPhard}, we have simplified it here considerably by restricting ourselves to local optimization and to limited time horizons. Both simplifications may imply a potentially reduced traffic performance in the urban street network, but this loss of performance is small if traffic lights adjust to arriving vehicle platoons \cite{jstat}. The reliable look-ahead times are anyway very limited for fundamental reasons (see the Appendix in Ref. \cite{jstat}). Therefore, one can restrict traffic light optimization to time periods $1/\lambda$, over which the traffic
forecast can be done with sufficient accuracy. When traffic lights are switched frequently, the value of $1/\lambda$ of the forecast time horizon will go down.
\par
Note that an optimization based on unreliable long-term forecasts will yield bad results. Therefore, it is not only {\it justified}, but also {\it successful} to replace the optimization of one or several full cycles by the optimization of, say, two phases. Alternatively, one may minimize the exponentially weighted travel times, i.e. minimize the function
\begin{equation}
 \widetilde{G} 
=  \sum_j \lambda I_j \int\limits_0^\infty \! dt \; \mbox{e}^{-\lambda t} \Big[ \Delta N_j(t) 
+  T'_j |C_j| \rho_{\rm jam} \Theta(\Delta N_j>0) \Big] 
\label{Goa}
\end{equation}
by variation of the duration and sequence of green phases.
While this approach is less suited for an analytical optimization, it reminds of formulations
of discounted functions in economics \cite{Feichtinger}. Goal function (\ref{Goa}) can be optimized  {\it numerically}, limiting the evaluation of the integral to the range $t<3/\gamma$. 

\end{document}